\documentclass[authoryear,3p,12pt]{elsarticle}

\usepackage[colorlinks=true]{hyperref}

\usepackage{graphics}
\usepackage{natbib}
\usepackage{setspace}
\usepackage{amssymb}
\usepackage{amstext}
\usepackage{textcmds}
\usepackage{color}
\usepackage{lineno}
\usepackage{longtable}
\usepackage{tabularx}
\usepackage{array}
\usepackage[version=3]{mhchem}

\small\normalsize

\journal{Icarus}

\begin{document}

\begin{frontmatter}

\title{2D photochemical modeling of Saturn's stratosphere. Part II: Feedback between composition and temperature}

\author[LAB1,LAB2,SwRI]{V. Hue\corref{cor1}} \ead{vhue$@$swri.org} 
\author[SwRI]{T. K. Greathouse}
\author[MPS,LESIA]{T. Cavali\'e}
\author[LAB1,LAB2]{M. Dobrijevic}
\author[LAB1,LAB2]{F. Hersant}

\address[LAB1]{Universit\'e de Bordeaux, Laboratoire d'Astrophysique de Bordeaux, UMR 5804, F-33270 Floirac, France}
\address[LAB2]{CNRS, Laboratoire d'Astrophysique de Bordeaux, UMR 5804, F-33270, Floirac, France}
\address[SwRI]{Southwest Research Institute, San Antonio, TX 78228, United States}
\address[MPS]{Max-Planck-Institut f\"ur Sonnensystemforschung, 37077, G\"ottingen, Germany}
\address[LESIA]{LESIA, Observatoire de Paris, CNRS, UPMC, Universit\'e Paris-Diderot, 5 place Jules Janssen, 92195 Meudon Cedex, France}


\date{Received March 13, 2015; accepted for publication in Icarus: December 08, 2015}

\begin{abstract}

Saturn's axial tilt of 26.7$^{\circ}$ produces seasons in a similar way as on Earth. Both the stratospheric temperature and composition are affected by this latitudinally varying insolation along Saturn's orbital path. The atmospheric thermal structure is controlled and regulated by the amount of hydrocarbons in the stratosphere, which act as absorbers and coolants from the UV to the far-IR spectral range, and this structure has an influence on the amount of hydrocarbons. We study here the feedback between the chemical composition and the thermal structure by coupling a latitudinal and seasonal photochemical model with a radiative seasonal model. Our results show that the seasonal temperature peak in the higher stratosphere, associated with the seasonal increase of insolation, is shifted earlier than the maximum insolation peak. This shift is increased with increasing latitudes and is caused by the low amount of stratospheric coolants in the spring season. At 80$^{\circ}$ in both hemispheres, the temperature peak at 10$^{-2}$\,mbar is seen to occur half a season (3-4 Earth years) earlier than was previously predicted by radiative seasonal models that assumed spatially and temporally uniform distribution of coolants. This shift progressively decreases with increasing pressure, up to around the 0.5\,mbar pressure level where it vanishes. On the opposite, the thermal field has a small feedback on the abundance distributions. Accounting for that feedback modifies the predicted equator-to-pole temperature gradient. The meridional gradients of temperature at the mbar pressure levels are better reproduced when this feedback is accounted for. At lower pressure levels, Saturn's stratospheric thermal structure seems to depart from pure radiative seasonal equilibrium as previously suggested by \citet{Guerlet2014}. Although the agreement with the absolute value of the stratospheric temperature observed by Cassini is moderate, it is a mandatory step toward a fully coupled GCM-photochemical model.


\end{abstract}

\begin{keyword}
\textbf{Jovian planets \sep Saturn, atmosphere \sep Atmospheres, structure \sep Photochemistry \sep Atmospheres, composition}
\end{keyword}

\end{frontmatter}

\section{Introduction}
\label{s:Model}

Similar to the Earth, Saturn's obliquity of 26.7$^{\circ}$ produces seasons forced by its 29.5 years orbital period. Insolation thus varies with latitude along that cycle. The atmospheric response to this forcing has been witnessed by Cassini spacecraft clearly in the stratospheric temperature and subtly in the chemical composition \citep{Fletcher2010,Orton2008,Sinclair2013}.

Saturn's stratospheric thermal structure is controlled by the heating due to near-IR methane (CH$_{4}$) absorption bands while the cooling is dominated by mid-IR emission lines of acetylene (C$_{2}$H$_{2}$), ethane (C$_{2}$H$_{6}$) and CH$_{4}$ \citep{Greathouse2008}. Saturn's stratospheric composition is, on the other hand, controlled by CH$_{4}$ photolysis in the UV spectral range. The recombination of the radicals produced by this photolysis initiates the production of numerous hydrocarbons, C$_{2}$H$_{6}$ and C$_{2}$H$_{2}$ being the most abundant ones (e.g. \citealt{Moses2005b, Hue2015}).

Consequently, there is an interesting feedback to study. The seasonally variable insolation received by Saturn as a function of latitude produces variations in the atmospheric heating/cooling rates, while producing, at the same time, variations in the amount of coolants due to photochemical processes. These processes have different timescales, meaning that only a 3D-GCM that would simultaneously solves the photochemistry, radiative and dynamical equations would be accurate. Such a solution requires computational means that are beyond our capabilities at the moment.

Therefore, the problem is usually decoupled using different numerical tools. The effect of the solar energy input on the Saturn's stratospheric temperature is inferred from radiative seasonal models (see e.g. \citealt{Bezard1985, Greathouse2008, Guerlet2014}) while its effect on the atmospheric composition is inferred from photochemical models \citep{Moses2005b, Dobrijevic2011,Hue2015}.

The already existing radiative seasonal models generally assume meridionally and temporally constant abundances \citep{Greathouse2008,Guerlet2014}. These assumed spatial distributions of coolants come from Cassini/CIRS observations \citep{Guerlet2009, Guerlet2010} or ground-based observations \citep{Greathouse2005}. However, the seasonal dependence of the spatial distribution of the coolants have been neglected so far. The main reason being that little temporal coverage of the observed distribution of hydrocarbons exists.

Until recently, the photochemical models published in the literature (e.g. \citealt{Moses2000a, Moses2005b}) did not account for the temporal evolution of temperature, or only tested the influence of its spatial variability as sensitivity cases. \citet{Hue2015} made a first attempt to use a realistic thermal field, i.e. a thermal field that evolves with latitude and time, in a pseudo-2D photochemical model. In that work, the stratospheric temperatures were computed by the radiative seasonal model of \citet{Greathouse2008}. Even in that study, the problem stayed partly uncoupled: the feedback between chemistry and temperature was not fully accounted for. In this paper, we study how the temperature is affected by using a more realistic distribution of atmospheric coolants and how such a temporally variable thermal field impacts the atmospheric composition in return. Furthermore, in the framework of the future development of more complex models that will solve at the same time the hydrodynamical and photochemical equations, we aim to assess if a seasonally repeatable state can be reached with such an approach.

In the first part of this paper, we describe how our radiative seasonal model and our photochemical model have been coupled. Then, we  present the feedback of each model on one another, i.e. the impact of a seasonally variable chemical composition on the predicted stratospheric temperatures and vice-versa. Finally, we compare the thermal field that accounts for that feedback with Cassini observations.

\section{Description of the model}
\label{s:Model}

We first briefly decribe here the photochemical model used to compute the chemical composition. Then, we describe the radiative seasonal model used to compute the stratospheric temperature. A more detailed decription of these models can be found in the associated publications.

\subsection{Photochemical model}
\label{ss:PhotochemicalModel}

The photochemical model used to compute the chemical abundances is the pseudo-2D (altitude-latitude) time-dependent model presented in \citet{Hue2015}. This model accounts for the variation of the seasonally variable parameters such as the subsolar latitude and the heliocentric distance. The ring shadow effects on the atmospheric chemistry are accounted for, following the prescription of \citet{Guerlet2014}, and are based on stellar occultation measurements published by \citet{Colwell2010}. The chemical scheme used in this model is based on the work of \citet{Loison2015}, in which the chemical scheme has been greatly improved thanks to previous work \citet{Hebrard2013} and \citet{Dobrijevic2014}. Following the methodology of \citet{Dobrijevic2011}, the chemical scheme has been reduced by \citet{Cavalie2015} in terms of number of species and reactions to make it useable by 2D-3D photochemical models of Saturn.


The photochemical model is divided into a spherical atmosphere consisting of 118 altitude levels and 17 latitude cells. The vertical grid span the pressure range of 10$^{3}$\,mbar to, at least, 10$^{-7}$\,mbar in order to fully resolve the depth of absorbed UV flux at the top of the model. The latitude grid ranges from 80$^{\circ}$S to 80$^{\circ}$N. The model solves the continuity equation using DLSODE from the ODEPACK library \citep{DLSODE}.

The photochemical equations were integrated over Saturn's sampled orbit. The hydrocarbon chemistry is mainly driven by methane photolysis that occurs in the higher stratosphere (around 10$^{-4}$ mbar). From there, the produced hydrocarbons diffuse down to the lower stratosphere. 

Finally, the pressure-temperature background used in the photochemical model comes from the radiative seasonal model described below.

\subsection{Radiative seasonal model}
\label{ss:SeasonalModel}

The temperature field is inferred from the multi-layer radiative seasonal model developed by \citet{Greathouse2008}. Consistently with the photochemical model presented above, it accounts for the same evolution in the seasonally variable parameters. The ring shadowing effects are accounted for following the formalism of \citet{Bezard1986} and \citet{Moses2005b}. The heating due to CH$_4$ absorption in the near-infrared, visible and UV spectral range is included as well as C$_2$H$_2$, C$_2$H$_4$ and C$_2$H$_6$ absorption in the UV range in a direct beam radiative transfer model (following the Beer-Lambert law). Cooling within the mid- to far-infrared range due to CH$_4$, C$_2$H$_2$ and C$_2$H$_6$ line emissions is included and follows the formalism of \citet{Goukenleuque2000}. The continuum emission in the far-infrared range due to collision-induced absorption of H$_2$-H$_2$, H$_2$-He and H$_2$-CH$_4$ is accounted for using the formalism of \citet{Borysow1985}, \citet{Borysow1986} and \citet{Borysow1988}.

The spectroscopic line information are taken from GEISA03 \citep{Geisa}, HITRAN \citep{Rothman2005} and from \citet{Vander2007}. From this information, $\kappa$-coefficient tables are produced for each molecule. This allows to modify each molecule abundance within a run independently from one another, while avoiding the re-calculation of the full $\kappa$-coefficient tables each time one of the absorbers/coolants abundance is altered. This substantially reduces the computational time.

The C$_2$H$_2$ and C$_2$H$_6$ distributions which were assumed for the first run come from Cassini/CIRS observations at planetographic latitude of 45$^{\circ}$S published by \citet{Guerlet2009}. These vertical profiles are kept constant with latitude and time to compute the initial thermal field.

The tropospheric aerosols are not accounted because the seasonal radiative model is focus here on predicting the stratospheric temperatures. The stratospheric aerosols have minor effects on the predicted temperatures and are not accounted here. The addition of the stratospheric aerosols following \citet{Karkoschka2005} increases temperatures by less than 1\,K above the 30\,mbar level at 83$^{\circ}$S. In the equatorial region, the temperatures are increased by less than 0.1\,K.

The radiative seasonal model extends from 9$\,\times\,$10$^{-5}$\,mbar to 660\,mbar. The model is divided into a plane parallel atmosphere consisting of 83 pressure levels. Local thermodynamic equilibrium (LTE) is assumed throughout. Temperatures are valid up to 10$^{-3}$\,mbar where the non-LTE effets start to dominate.

\section{Thermal and chemical feedbacks}
\label{s:Chemical_feedback}



We now describe how the two models have been iterated forming a pseudo-coupling. First, we run the radiative seasonal model assuming the uniform distribution of absorbers and coolants retrieved by \citet{Guerlet2009}. Then, we used the resulting thermal field as input for the photochemical model. The seasonal distribution of abundances obtained at this point of the study are identical to those presented by \citet{Hue2015}, i.e. their (S) thermal field study case. Then, the radiative seasonal model was run again accounting for the seasonal distribution of absorbers and coolants obtained from the photochemical model.

Each time a new thermal field is produced and then used as input parameters for the photochemical model (i.e. at every iteration), the latter model was run over 30 orbits for the abundances to converge down to the 1\,bar pressure level, following the methodology of \citet{Hue2015}. The 30 orbits required to reach the seasonal convergence are due to the way the photochemical model was initially built. Indeed, every photochemical simulation starts from an atmosphere mostly composed of the 3 main compounds (H$_2$, He, CH$_4$). From that point, the photochemistry produces hydrocarbons, which then diffuse toward the lower stratosphere. Similarly, each time a new abundance field is obtained, the radiative seasonal model is also run over 3 orbits for the predicted thermal field to reach a seasonal convergence. The 3 orbits are needed due to the fact that the seasonal radiative model also starts from a standard thermal profile, based on observations.

We made sure that running both models until seasonal convergence at every iteration do not affect the overall conclusions. Therefore, we checked that we obtain the same results by running both the seasonal photochemical model and the radiative seasonal model over 1 orbit at each iteration. In particular, instead of starting the radiative and photochemical calculations from a standard condition, i.e. standard thermal profile for the radiative model and standard chemical composition for the photochemical model, both models were started from the conditions obtained at the previous iteration. When proceeding this way, as soon as half the Saturn's year, the computed thermal field and abundance distribution become identical to the ones predicted when both models are run until seasonal convergence.

\begin{figure}[htp]
   \centering
   \includegraphics[width=0.8\columnwidth]{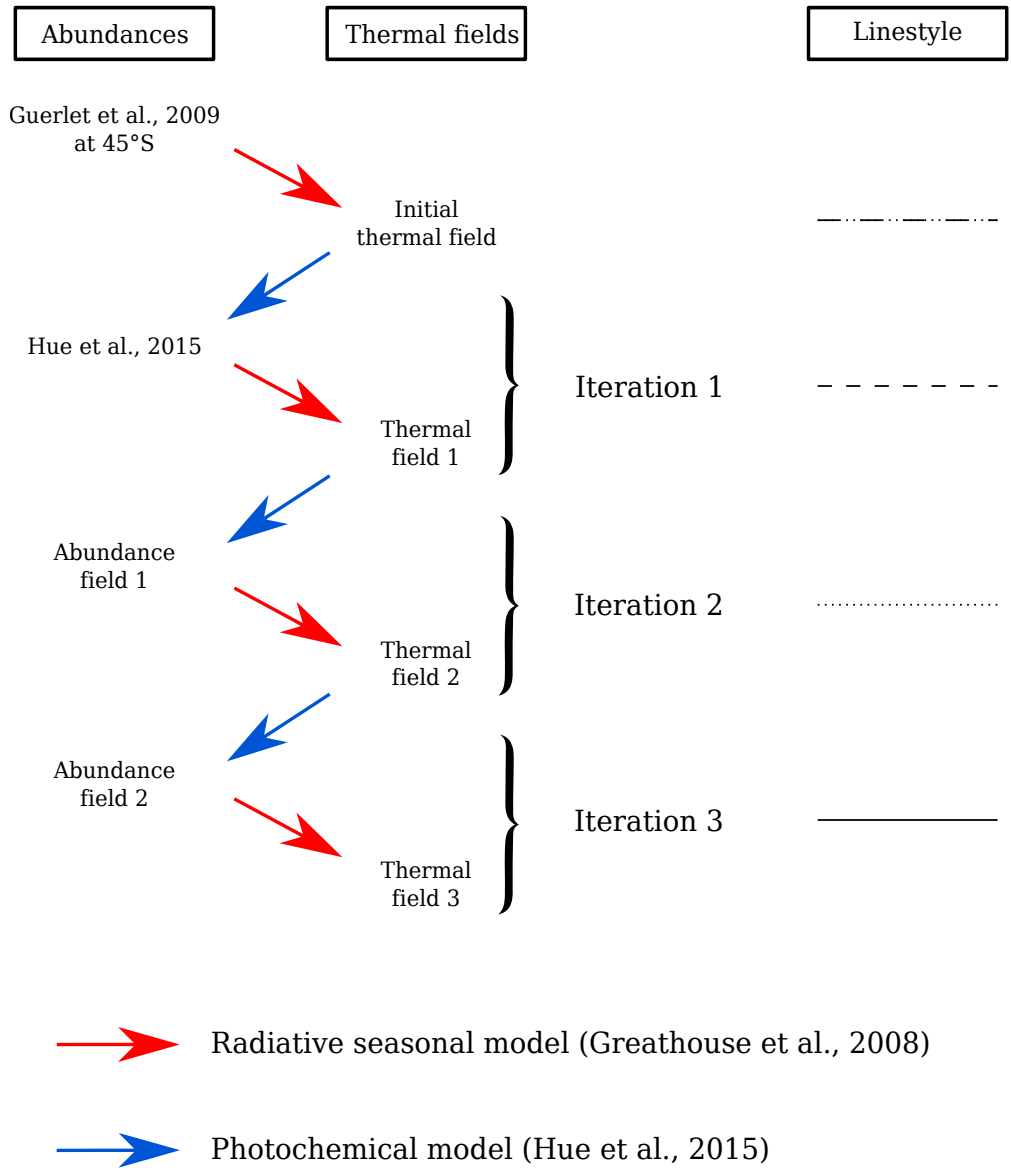}
      \caption{Presentation of the coupling between the radiative seasonal model of \citet{Greathouse2008} and the photochemical model of \citet{Hue2015}. We used the linestyle associated with each iteration presented in this figure for the following plots. The use of the initial thermal field in the photochemical model reproduces the results of the (S) thermal field of \citet{Hue2015}. Red arrows denote the runs of the radiative seasonal model with a given seasonally varying abundance distribution (except for the initial thermal field model) while blue arrows denote the runs of the photochemical model with a given seasonally variable thermal field.
      }
      \label{fig:Chart}
   \end{figure}

A chart showing the way the two models have been pseudo-coupled is displayed in Fig. \ref{fig:Chart}. The linestyles used in the following plots are presented in this figure.

These steps were repeated until the convergence between the two models was reached, i.e. when the differences in the temperature from one iteration to the other were below 0.5\,K at all pressure levels. The feedback of the thermal field on the abundance fields is weaker than the feedback of the abundance fields on the thermal field. Therefore, the threshold for which the iterations are stopped is only applied to the thermal field.

\citet{Greathouse2008} showed that, depending on the stratospheric level considered, the chemical species that is responsible for the main cooling effect is expected to vary. For instance, they showed that C$_2$H$_2$ is the main coolant at 10$^{-2}$\,mbar, while C$_2$H$_6$ generally dominates the cooling at 0.1\,mbar. However, all the hydrocarbons, C$_2$H$_2$, C$_2$H$_6$ and CH$_4$, contribute to the cooling from 10\,mbar to the top of the atmosphere. In the lower stratosphere, at the 10\,mbar level and below, the collision-induced absorption of H$_2$-H$_2$, H$_2$-He and H$_2$-CH$_4$ is predicted to dominate the cooling with a minor addition by C$_2$H$_6$.

We present the seasonal evolution of the temperature at the different iterations, as well as the C$_2$H$_2$ and C$_2$H$_6$ abundances at 10$^{-2}$\,mbar (Fig. \ref{fig:Results_1}) and 1\,mbar (Fig. \ref{fig:Results_3}) where they dominate the cooling. \citet{Hue2015} showed that the impact of a seasonally variable thermal field on the chemical composition was maximum where the seasonal thermal gradients were also maximum, i.e. the higher the latitude, the stronger these effects. Therefore, we will only present the results at 80$^{\circ}$S and 40$^{\circ}$N, where the feedback between the chemical composition and stratospheric temperature is expected to be strong. The same general trends are observed at both latitudes, although more pronounced at 80$^{\circ}$S. Therefore, in the following sections, we focus mainly on discussing the results at 80$^{\circ}$S. For the sake of comprehension, the linestyles have been kept coherent between temperatures and hydrocarbon abundances (see Fig. \ref{fig:Chart}).



\subsection{Thermal and chemical feedbacks in the upper stratosphere (10$^{-2}$\,mbar)}

\begin{figure}[htp]
   \centering
   \includegraphics[width=0.7\columnwidth]{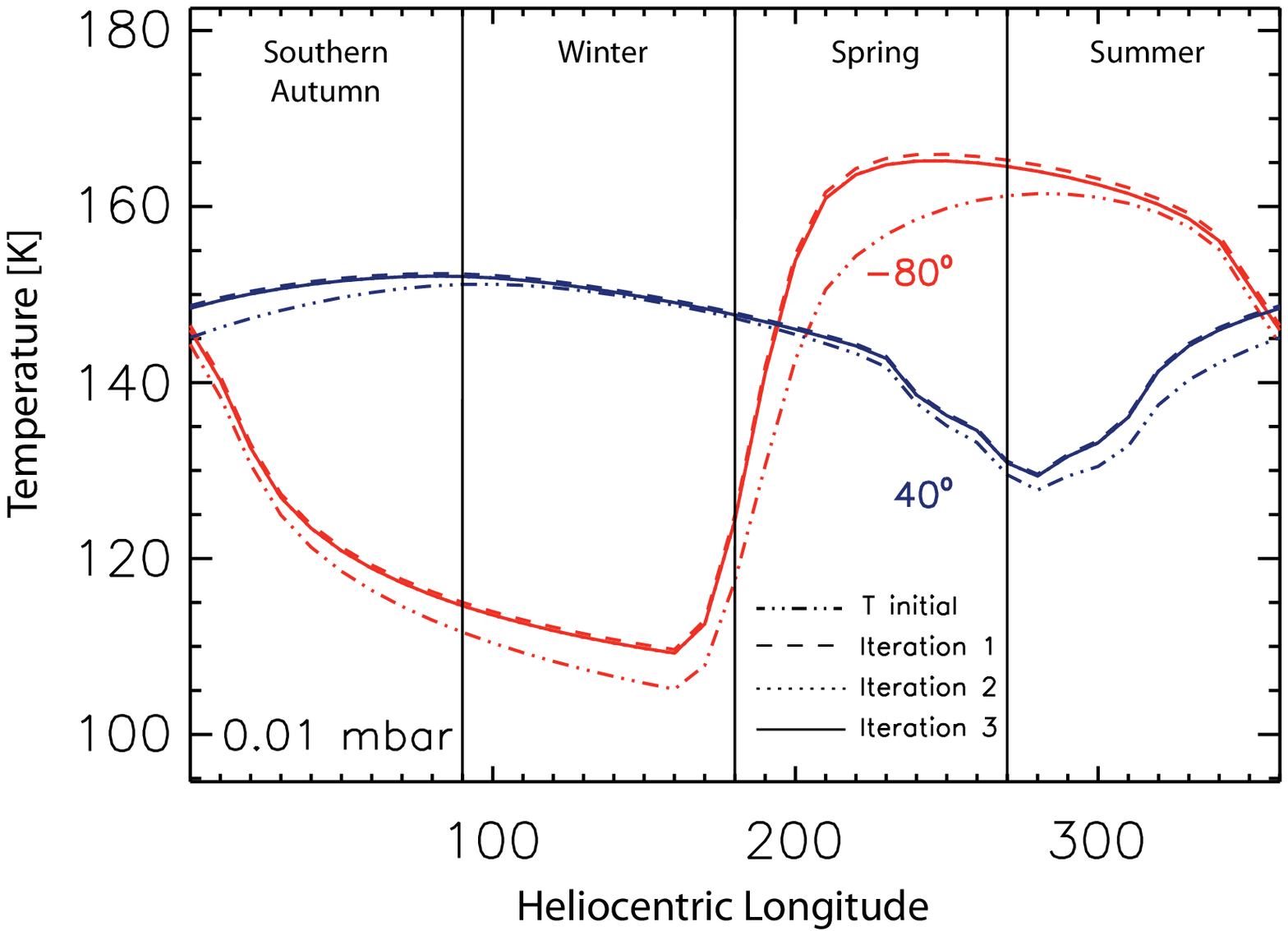}\\
   \includegraphics[width=0.7\columnwidth]{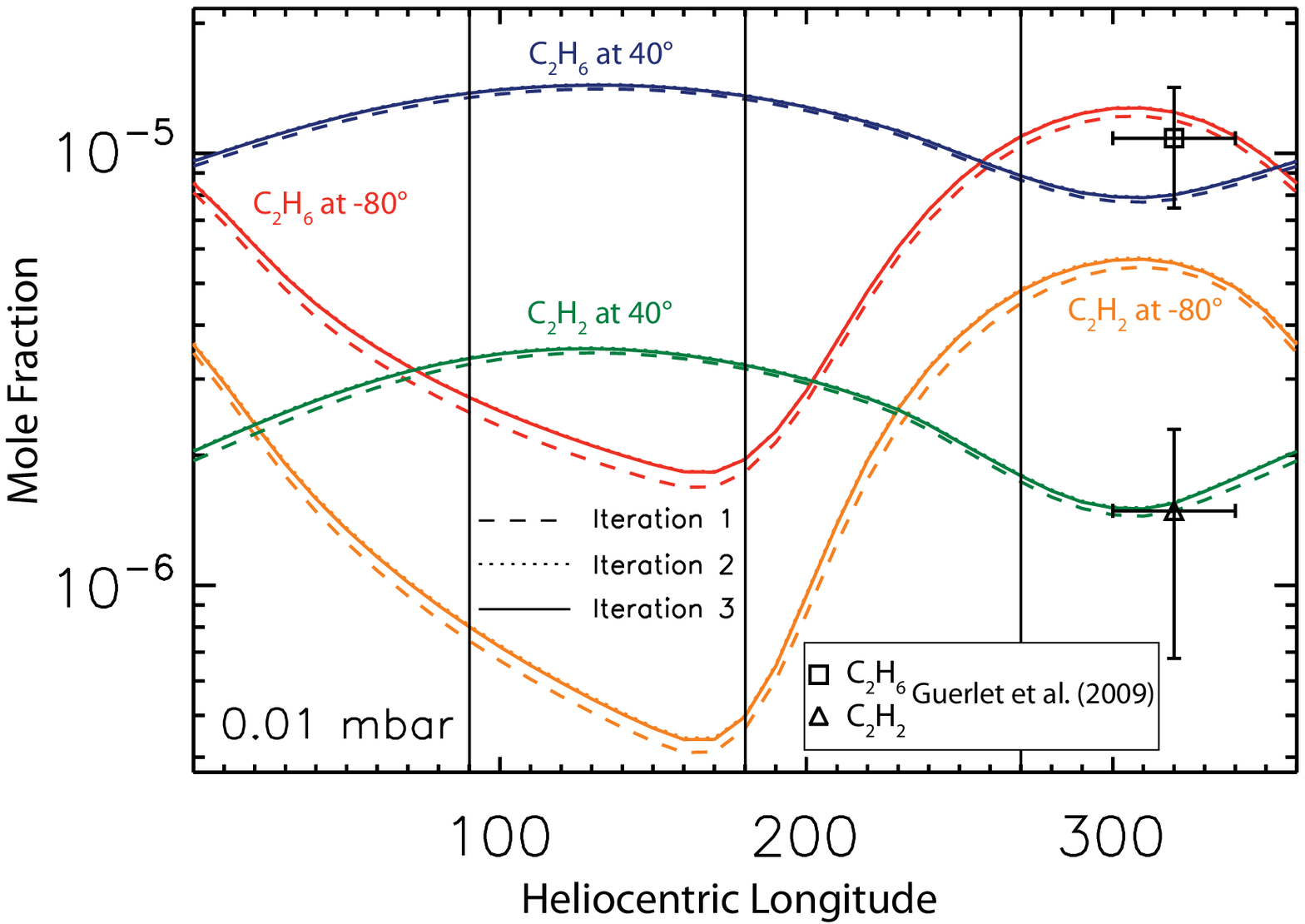}
      \caption{Seasonal evolution of temperature (upper panel) and hydrocarbon abundances (C$_2$H$_2$ and C$_2$H$_6$, lower panel) at 10$^{-2}$\,mbar and planetocentric latitudes of 40$^{\circ}$N and 80$^{\circ}$S. The various linetypes depict the different iterations between the photochemical and radiative seasonal models. At 40$^{\circ}$N, C$_2$H$_2$ is displayed in green and C$_2$H$_6$ in blue. At 80$^{\circ}$S, C$_2$H$_2$ is displayed in orange and C$_2$H$_6$ in red. The square and triangle data points represent respectively the C$_{2}$H$_{6}$ and C$_{2}$H$_{2}$ hydrocarbon abundances observed by Cassini \citep{Guerlet2009} at planetographic latitude of 45$^{\circ}$S and which were used to compute the initial thermal field. These initial hydrocarbon distributions are assumed to be constant with latitude and time during the first run of the radiative seasonal model. The horizontal uncertainty on Cassini/CIRS observations comes from the temporal range over which these observations have been performed.
      }
      \label{fig:Results_1}
   \end{figure}

At 10$^{-2}$\,mbar (Fig. \ref{fig:Results_1}), the initial run of the radiative seasonal model predicts that the temperature (upper panel) reaches a maximum shortly after the summer solstice. Since this thermal field accounts for a latitudinally uniform distribution of coolants, the solar zenith angle and Saturn's heliocentric distance are the main quantities driving the temporal stratospheric temperature peak at this pressure level.

After the first iteration, we noticed important changes in the thermal field at 80$^{\circ}$S, up to 12\,K for L$_s$ = 200$^{\circ}$, because the radiative seasonal model now assumes a seasonally variable distribution of coolants. The seasonal temperature peak at 10$^{-2}$\,mbar is now occurring before the southern solstice. The origin of this early temperature peak is caused by the small amount of C$_2$H$_2$ and C$_2$H$_6$ during the early spring season in the southern hemisphere from mid- to high-latitudes, as predicted by the photochemical model (lower panel on Fig. \ref{fig:Results_1}). Indeed, the photochemical model predicts that the amounts of C$_2$H$_2$ and C$_2$H$_6$ are maximum after the summer solstice. Therefore, in the early spring season, the increase in the upper-stratospheric temperature, mainly due to CH$_4$ near-IR absorption is not efficiently compensated by the mid- to far-IR cooling of C$_2$H$_2$ and C$_2$H$_6$, whose abundances are low at that time of the year. Later during that season, when the CH$_4$ UV-photolysis has led to the production of substantial amounts of C$_2$H$_2$ and C$_2$H$_6$, the temperature at this level shows the opposite trend and starts decreasing from mid-spring until the summer season in the southern hemisphere. Similarly, during the autumn and winter seasons, the predicted amount of C$_2$H$_2$ and C$_2$H$_6$ at iteration 1 are now lower than in the initial one. Due to this lower amount of coolants during that period (L$_s$ = 0$^{\circ}$-180$^{\circ}$), the predicted thermal field is hotter than the one using the initial distribution of coolants observed by Cassini \citep{Guerlet2009}.

The seasonal evolution of the thermal field at the following iterations are very similar to the one predicted at iteration 1, except during the spring season where little differences are noted. At 10$^{-2}$\,mbar, the differences between iteration 2 and 1 at 80$^{\circ}$S are up to 0.72\,K for L$_s$ = 260$^{\circ}$. The convergence is reached after 3 iterations with the photochemical model.

\begin{figure}[htp]
   \centering
   \includegraphics[width=0.6\columnwidth]{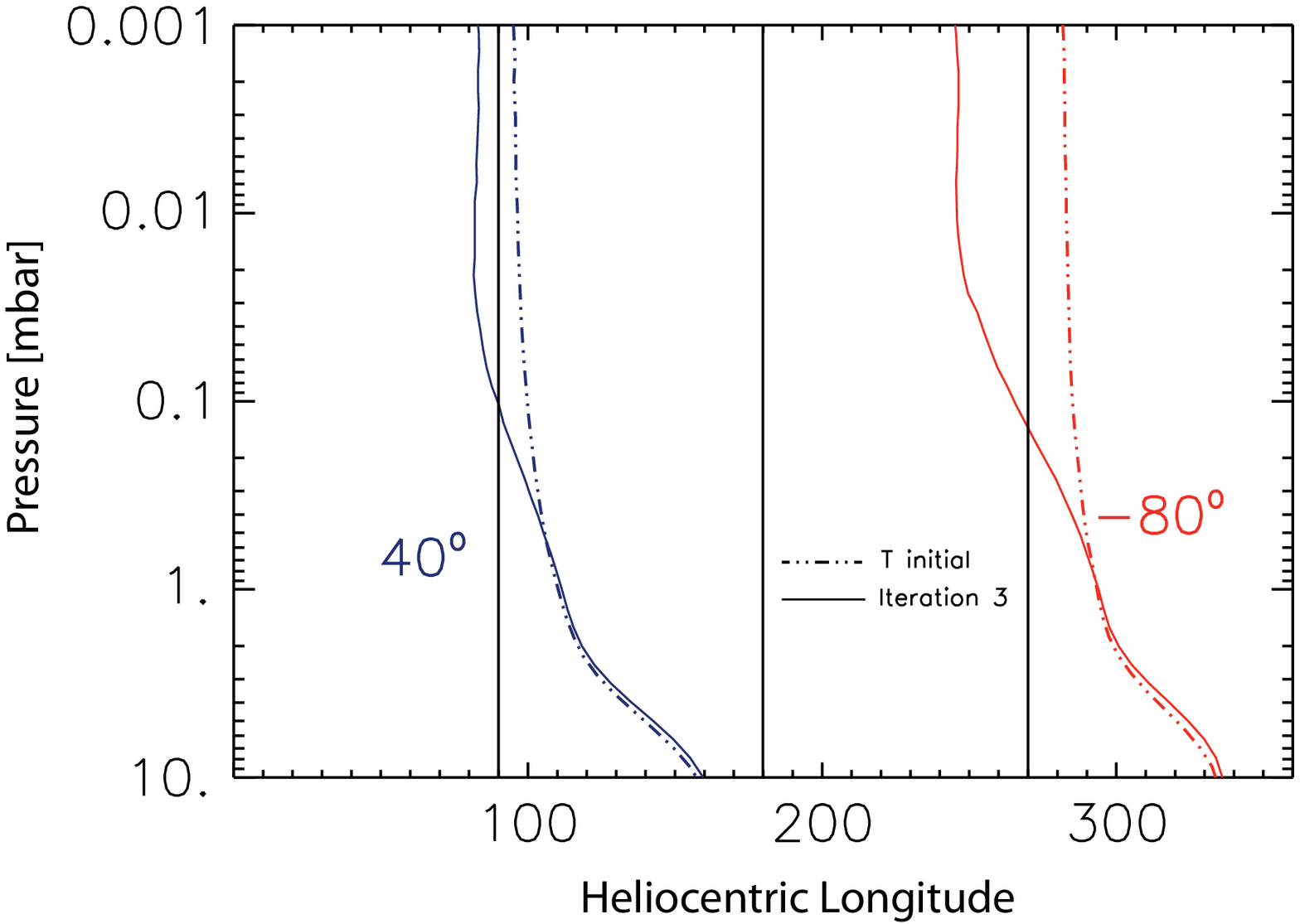}
      \caption{Evolution of the heliocentric longitude of the seasonal temperature peak as a function of pressure. The initial thermal fields and the converged thermal field are displayed by the dashed-double dotted lines and the solid lines, respectively. Planetographic latitudes of 40$^{\circ}$N and 80$^{\circ}$S are represented in blue and red, respectively.
      }
      \label{fig:Temperature_Shift}
   \end{figure}

The moment of the year when the initial thermal field and the converged thermal field reach a maximum is displayed in Fig. \ref{fig:Temperature_Shift}. At 80$^{\circ}$S, the seasonal temperature peak is shifted with respect to the maximum insolation (which occurs at L$_S$ = 270$^{\circ}$ in the southern hemisphere) due to Saturn's perihelion. Indeed, at 10$^{-2}$\,mbar, the temperature peak occured at L$_S$ = 280$^{\circ}$ in the initial model and was found to shift to L$_S$ = 243$^{\circ}$ at iteration 3. At this latitude and pressure level, this shift of -37$^{\circ}$ in L$_S$ corresponds to the peak occurring 2.8 Earth years earlier. At 80$^{\circ}$N, the situation is slightly different, because of Saturn's eccentric orbit. Indeed, at that latitude and season, the temperature peak at 10$^{-2}$\,mbar is shifted 47$^{\circ}$ earlier in L$_S$, which corresponds to 4.2 Earth years.

The modification of the thermal field affects the seasonal distribution of C$_2$H$_2$ and C$_2$H$_6$, as shown by \citet{Hue2015}. First, the modification of the thermal field amplitude will affect the location of the homopause, by shifting it to lower-pressure levels if the thermal field becomes hotter than the initial one. This shift in the position of the homopause will therefore modify the integrated production rates of radicals, which subsequently affects the production of hydrocarbons. Then, the increase in the seasonal thermal gradient will affect the diffusion of the photolysis by-products to higher-pressure levels, throughout a greater seasonal contraction/dilation of the atmospheric columns. Indeed, \citet{Hue2015} showed that, when the atmosphere contracts/expands during winter/summer time, the vertical mole fraction gradient on the altitude grid increases/decreases and causes the compounds to diffuse faster/slower toward the lower atmosphere.

This last feedback on the predicted distribution of hydrocarbons is observed during the southern winter season although that effect is not significant when compared to the error bars on the Cassini CIRS limb observations \citep{Guerlet2009}. The predicted temperatures of iteration 1 are greater than the initial ones during the winter season, therefore the contraction of the atmospheric column during that season is less important. This therefore causes the predicted abundances of C$_2$H$_2$ and C$_2$H$_6$ to remain higher through iterations 2 and 3.


At 10$^{-2}$\,mbar, the feedback in the C$_2$H$_2$ abundance field between the converged and initial thermal field never exceeds 11\%. The maximum value of that feedback is occurring during the early southern spring at 80$^{\circ}$S. That feedback is even weaker for C$_2$H$_6$, which never exceeds 9\%.


An identical behaviour occurs at 40$^{\circ}$N (Fig. \ref{fig:Results_1}, blue and green lines). At this latitude, the thermal and chemical feedbacks are similar to the one observed at 80$^{\circ}$S, although less pronounced due to the lower seasonal variation of insolation. The southern early spring temperature peak is also noted as well as the subsequent feedback between chemical composition and atmospheric temperature through the following iterations.

\begin{figure}[htp]
   \centering
   \includegraphics[width=0.7\columnwidth]{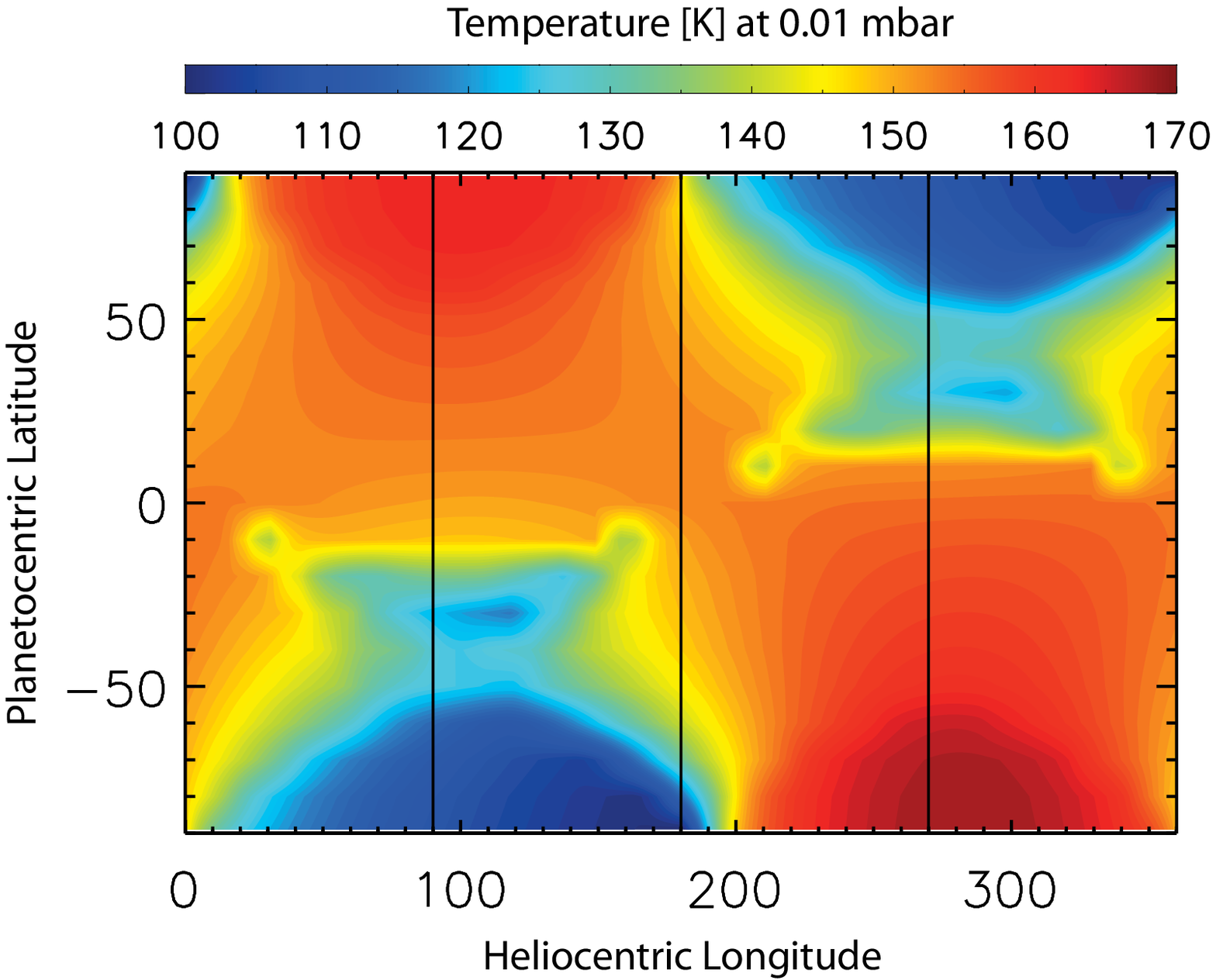}\\
   \includegraphics[width=0.7\columnwidth]{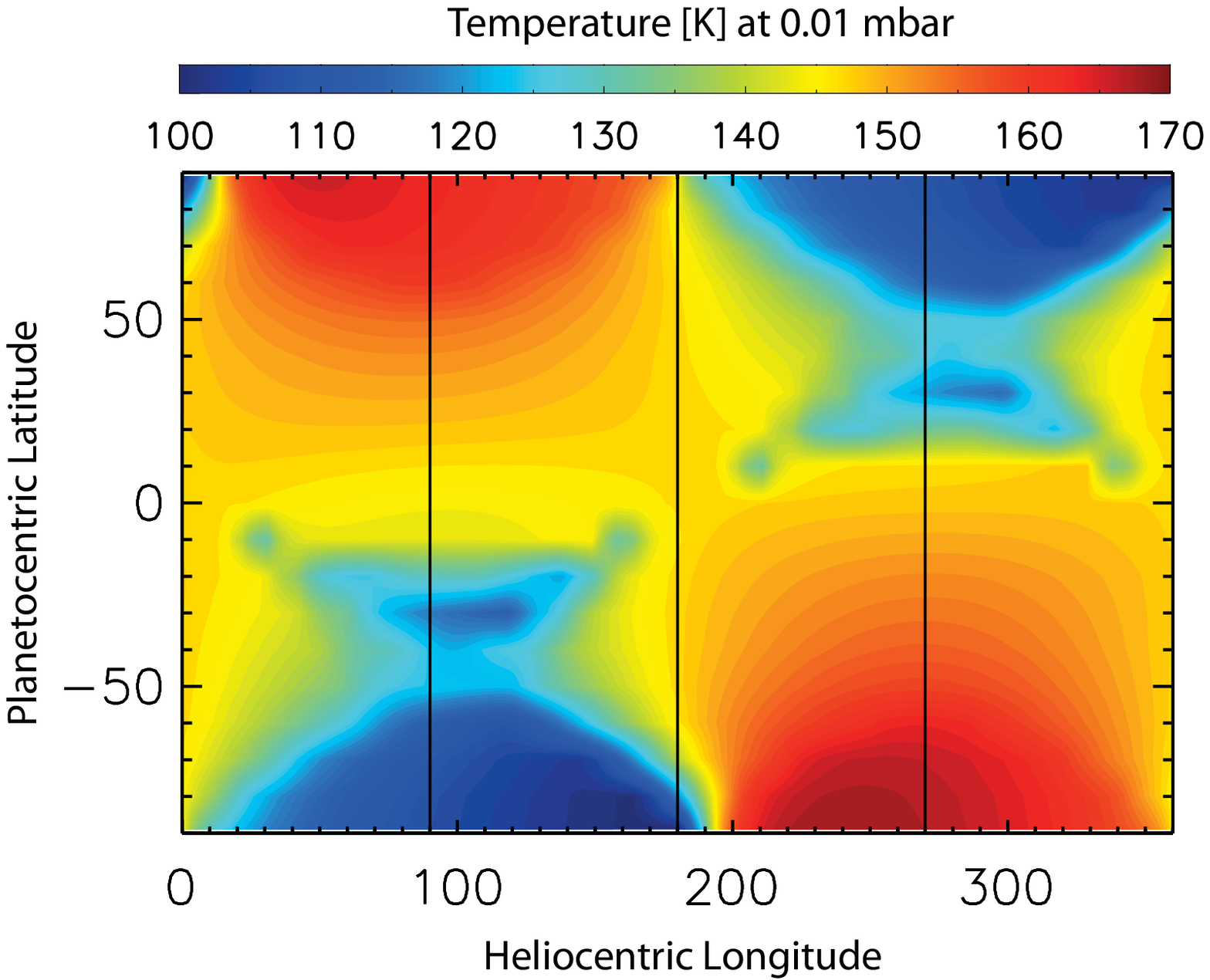}
      \caption{Thermal field at 10$^{-2}$\,mbar as a function of planetocentric latitude and heliocentric longitude. The upper panel depicts the initial thermal field that assumes a spatio-temporally uniform distribution of C$_2$H$_2$ and C$_2$H$_6$ observed by Cassini/CIRS \citep{Guerlet2009}. The lower panel presents the converged thermal field, after 3 iterations with the photochemical model.}
      \label{fig:2D_field_converged}
   \end{figure}

The seasonal evolution of the initial thermal field as well as the converged thermal field (i.e. iteration 3) at 10$^{-2}$\,mbar are presented in Fig. \ref{fig:2D_field_converged}. The blue feature at mid-latitudes in the winter hemisphere, centered around the winter solstice in this hemisphere, is caused by the ring shadowing effect (see for instance \citet{Moses2005b} and \citet{Hue2015}). The time-shift in heliocentric longitude (or L$_s$) of the temperature peak in the spring season increases with increasing latitude though it is also dependent of the eccentricity of Saturn giving the north a slightly exaggerated shift relative to similar southern latitudes.

At 10$^{-2}$\,mbar, the feedback between chemical composition and stratospheric temperature is important, because C$_2$H$_2$ and C$_2$H$_6$ abundances still show significant seasonal variability. However, this situation is no longer true at higher-pressure levels, as recent photochemical models \citep{Moses2005b,Hue2015} have shown that the seasonal variability decreases with increasing pressure. This decrease in the seasonal variability is due to the fact that these molecules, and more especially C$_2$H$_6$, are primarily diffusing to greater depth in the atmosphere, and are thus controlled by diffusion rather than photochemical production and loss.


\subsection{Thermal and chemical feedbacks in the lower stratosphere (1\,mbar)}

\begin{figure}[htp]
   \centering
   \includegraphics[width=0.7\columnwidth]{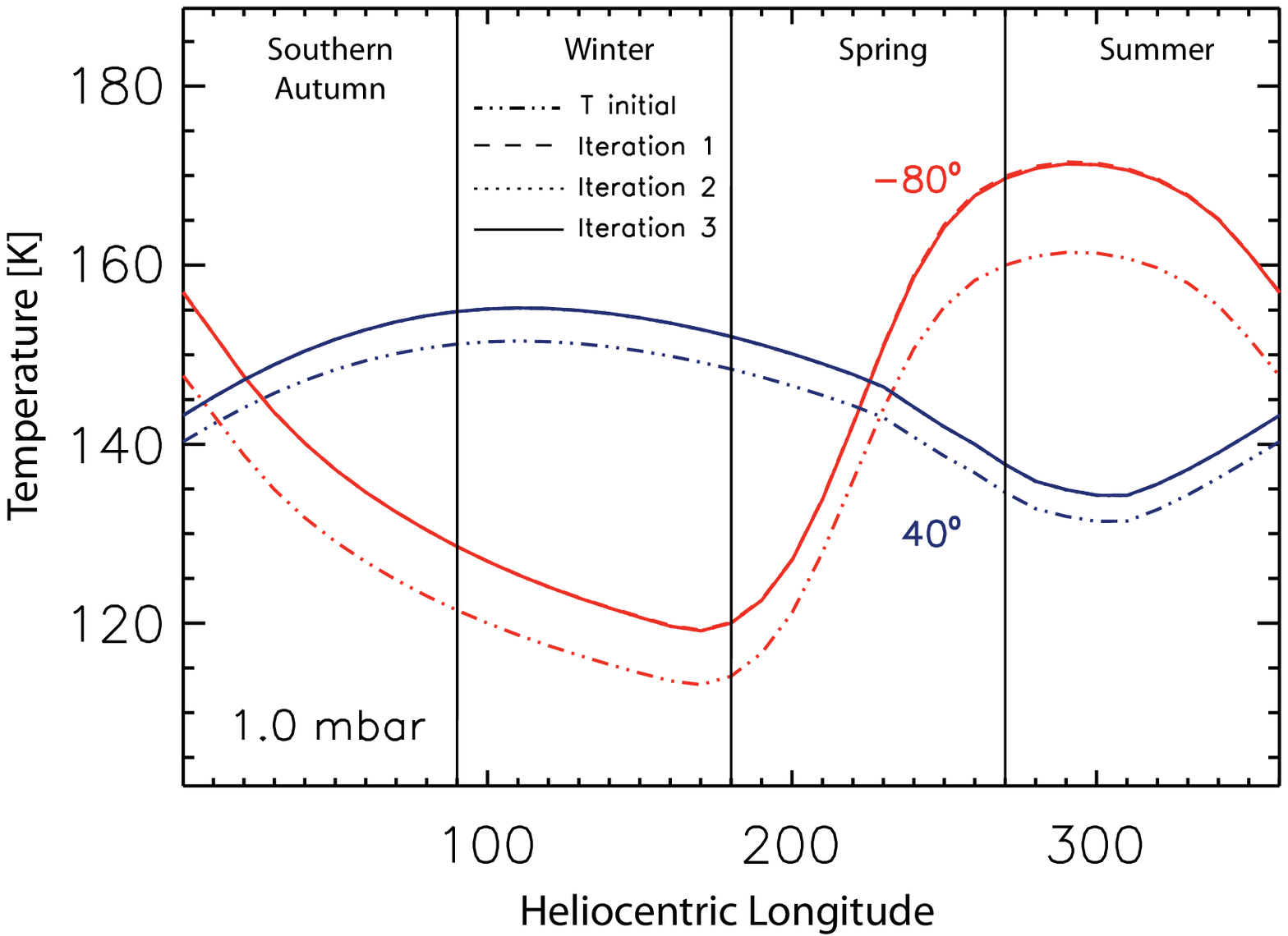}\\
   \includegraphics[width=0.7\columnwidth]{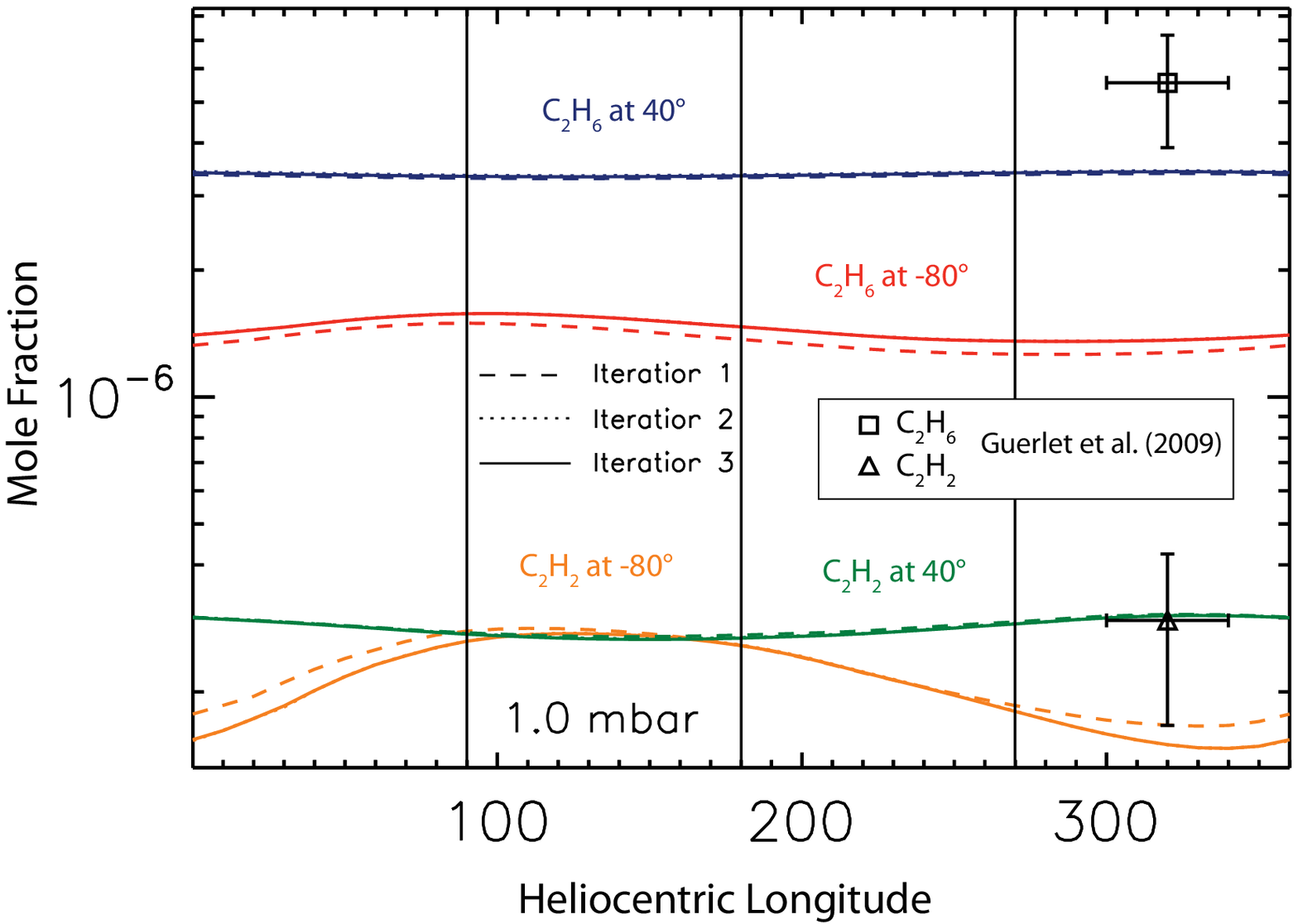}
      \caption{Same as Fig. \ref{fig:Results_1} for pressure level of 1\,mbar. The square and triangle data points represent respectively the C$_{2}$H$_{6}$ and C$_{2}$H$_{2}$ hydrocarbon abundances observed by Cassini \citep{Guerlet2009} and which were used to compute the initial thermal field.}
      \label{fig:Results_3}
   \end{figure}

At 1\,mbar (Fig. \ref{fig:Results_3}), the feedback between chemical composition and stratospheric temperature is less pronounced than it is in the upper stratosphere. The shift in time at which the summer temperature peak occurs in the converged thermal field relative to the initial thermal field decreases with increasing pressure through the stratosphere (Fig. \ref{fig:Temperature_Shift}).

At 1\,mbar, the summer temperature peak of the converged thermal field is occurring shortly after the summer solstice, similar to the behavior of the initial thermal field. The seasonal variability of C$_2$H$_6$ is small while it is still significant for C$_2$H$_2$, in agreement with \citet{Moses2005b} and \citet{Hue2015}. The abundances of C$_2$H$_6$ and C$_2$H$_2$ now show a seasonal variation of 16\% and 87\%, respectively. C$_2$H$_6$ is the main coolant at 1\,mbar for most of the year except at high latitudes during the winter season, where the collision-induced far-IR opacities dominate the cooling \citep{Greathouse2008}. Thus, the thermal feedback that results from accounting for a seasonally variable atmospheric composition with respect to a supposed constant hydrocarbon distribution is weak here, due to the predicted low seasonal variability of C$_2$H$_6$.

\begin{figure}[htp]
   \centering
   \includegraphics[width=0.6\columnwidth]{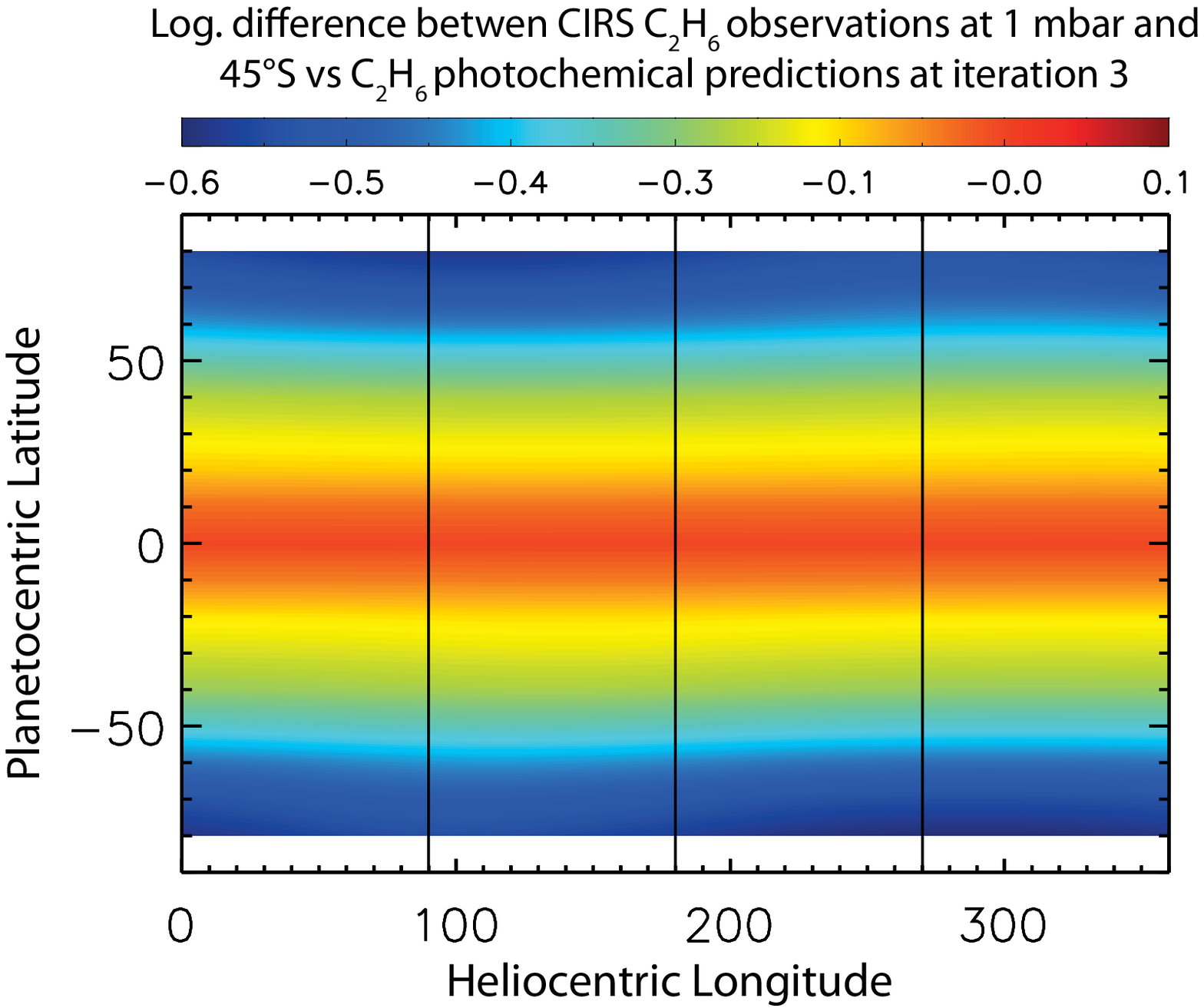}
      \caption{Logarithmic differences between the C$_2$H$_6$ abundances predicted at iteration 3 with the Cassini observations at planetographic latitude of 45$^{\circ}$S recorded at L$_s$ = 320$^{\circ}$ that were used as a priori to compute the initial thermal field. Red/blue colors denote an over/underprediction of the C$_2$H$_6$ abundance with respect to the Cassini observations.
      }
      \label{fig:Diff_Log}
   \end{figure}

Inter-comparisons of the C$_2$H$_6$ abundance assumed in the radiative seasonal model from the initial thermal field to the converged field are linked to the differences in the predicted temperatures between these two iterations. The modeled C$_2$H$_6$ abundances from mid- to polar latitudes are underpredicted at the 1 mbar pressure level with respect to the Cassini observations that were used as the a priori to compute the initial thermal field. This underprediction increases with latitude toward the poles. The logarithmic difference between the C$_2$H$_6$ abundances computed at iteration 3 with the Cassini observations at planetographic latitude of 45$^{\circ}$S is displayed in Fig. \ref{fig:Diff_Log}. At iteration 3, the C$_2$H$_6$ equatorial abundances are overestimated by a factor of 1.3. These abundances are gradually underestimated with latitude to reach a factor of 4 at $\pm$ 80$^{\circ}$. The predicted thermal fields at iterations 1 to 3 are subsequently hotter than the initial one.

C$_2$H$_6$ is mainly produced by the CH$_3$-CH$_3$ reaction around 10$^{-4}$\,mbar and then diffuses to higher-pressure levels. The seasonal evolution of C$_2$H$_6$ at pressures larger than 10$^{-4}$\,mbar is therefore related to the seasonal evolution of C$_2$H$_6$ at 10$^{-4}$\,mbar with a phase lag due to the slower mixing of the C$_2$H$_6$ down to greater pressures. A clear correlation is observed between the seasonal evolution of C$_2$H$_6$ at 10$^{-2}$\,mbar and 1$\,$mbar, between the different iterations at 80$^{\circ}$S. The increase noted at 1$\,$mbar peaked at L$_S$ = 110$^{\circ}$ in the C$_2$H$_6$ abundance at iteration 3 corresponds to the predicted C$_2$H$_6$ abundance peak at L$_S$ = 310$^{\circ}$ of the previous year at 10$^{-2}$\,mbar. At 1\,mbar, the feedback in the C$_2$H$_6$ abundance field between the converged and the initial thermal field reaches 7\% in the early spring season.

On the other hand, C$_2$H$_2$ main production regions are deeper in the atmosphere than those for C$_2$H$_6$. Indeed, C$_2$H$_2$ is mainly produced around 10$^{-2}$\,mbar and around 1\,mbar while C$_2$H$_6$ has a major production region around 10$^{-4}$\,mbar.
At the latter two levels, the impact of the thermal field amplitude on the reaction rates is negligible. Therefore, the influence of the thermal field on the C$_2$H$_2$ 1\,mbar abundance is mainly caused by variations in the diffusion rates due to the changing in the scale height of the atmosphere throughout a season. C$_2$H$_2$ is more sensitive to the increase in the thermal field variability through the faster diffusion because its vertical gradient is higher than that of C$_2$H$_6$. At 1\,mbar, the feedback in the C$_2$H$_2$ abundance field between the converged and the initial thermal field reaches 13\% around the southern autumn equinox.


\section{Comparison with Cassini observations}
\label{s:Comparison}

Thanks to the Cassini mission, observations of Saturn's stratospheric temperature and hydrocarbon emissions have been performed with unprecedented spatial and temporal coverage, either in nadir \citep{Sinclair2013, Fletcher2010} or limb \citep{Guerlet2009, Guerlet2010, Guerlet2011, Sylvestre2015} observing geometries. The thermal field has a small feedback on the abundance distribution (refer to the bottom plots of Figs. \ref{fig:Results_1} and \ref{fig:Results_3}), i.e. the abundance changes as a function of iteration are not significant when compared to the observation error bars. The most important changes on the abundance fields occur at iteration 1, which was previously compared to Cassini observations in \citet{Hue2015}. In that previous work, the C$_2$H$_2$ and C$_2$H$_6$ predicted abundances reproduced well the Cassini observations from the equator to mid-latitudes at pressures greater than 0.1\,mbar. In the southern hemisphere, these molecule abundances were underpredicted from 40$^{\circ}$S to south pole at pressures greater than 0.1\,mbar, and overpredicted at latitudes ranging from 10$^{\circ}$S to 40$^{\circ}$S at pressures lower than 0.1\,mbar. 
In that paper, these under/overprediction regions were interpreted in terms of a large-scale stratospheric circulation cell which is thought to move away the compound abundances from pure photochemical predictions. We refer the reader to the work of \citet{Hue2015} for a more detailed discussion on that subject. Therefore, we only focused on how the converged thermal field reproduced the observations of temperature.

We present now a comparison between the initial and converged thermal fields with Cassini observations of the stratospheric temperature both in nadir and limb observing geometries.

\subsection{Comparison with nadir-observations}
\label{sss:Limb_Observations}

\begin{figure}[htp]
   \centering
   \includegraphics[width=0.55\columnwidth]{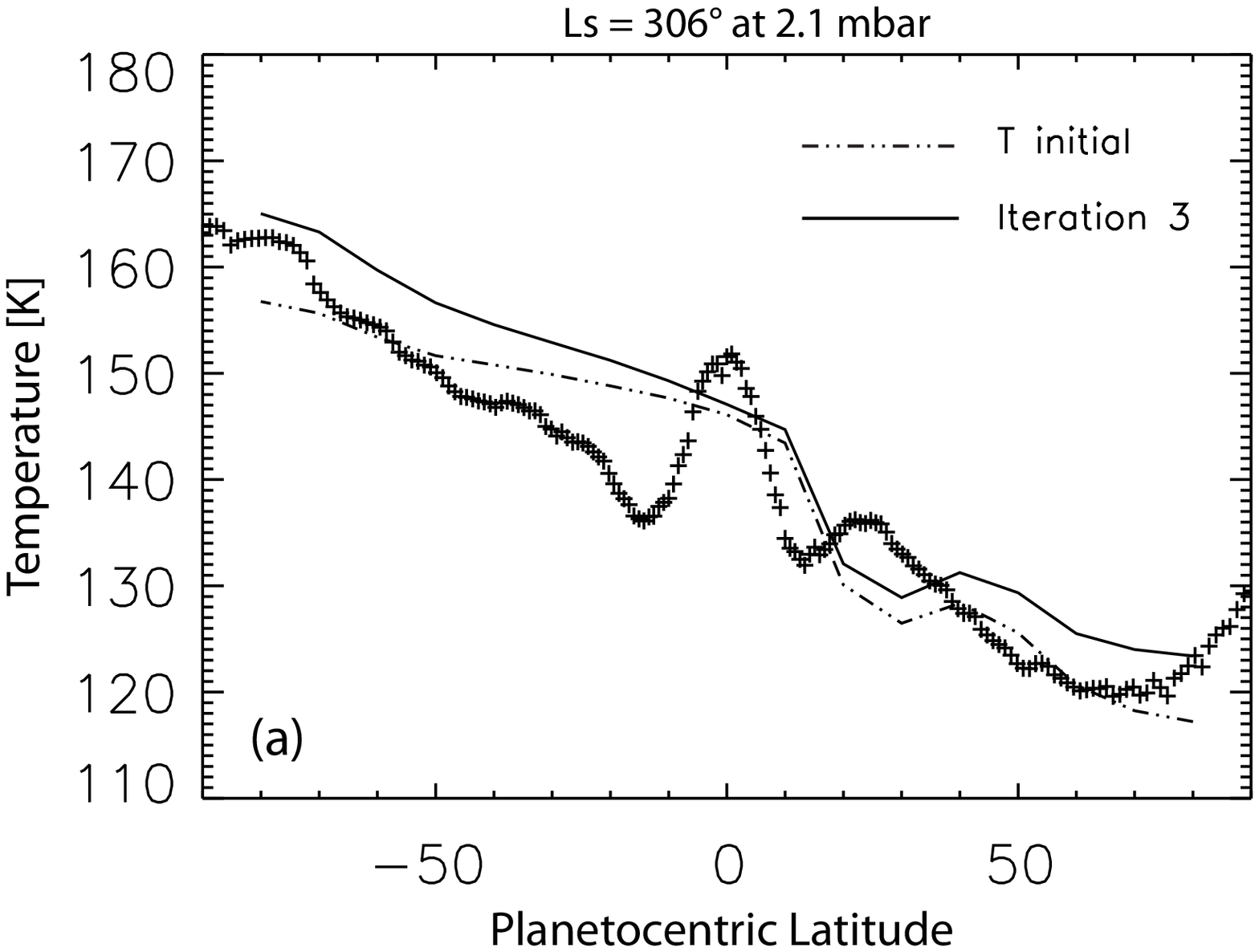}\vspace{0.3cm} \\
   \includegraphics[width=0.55\columnwidth]{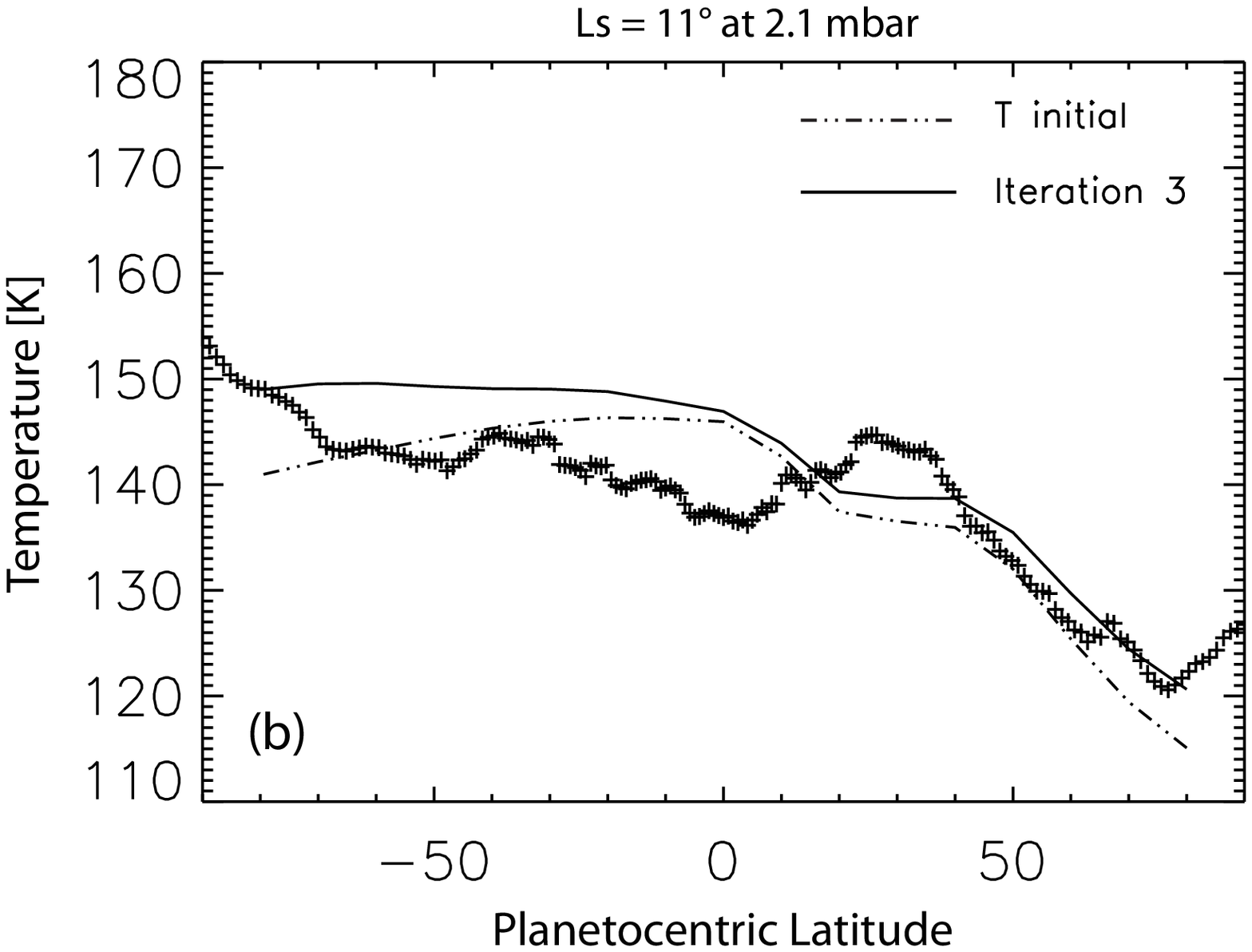}\vspace{0.3cm} \\
   \includegraphics[width=0.55\columnwidth]{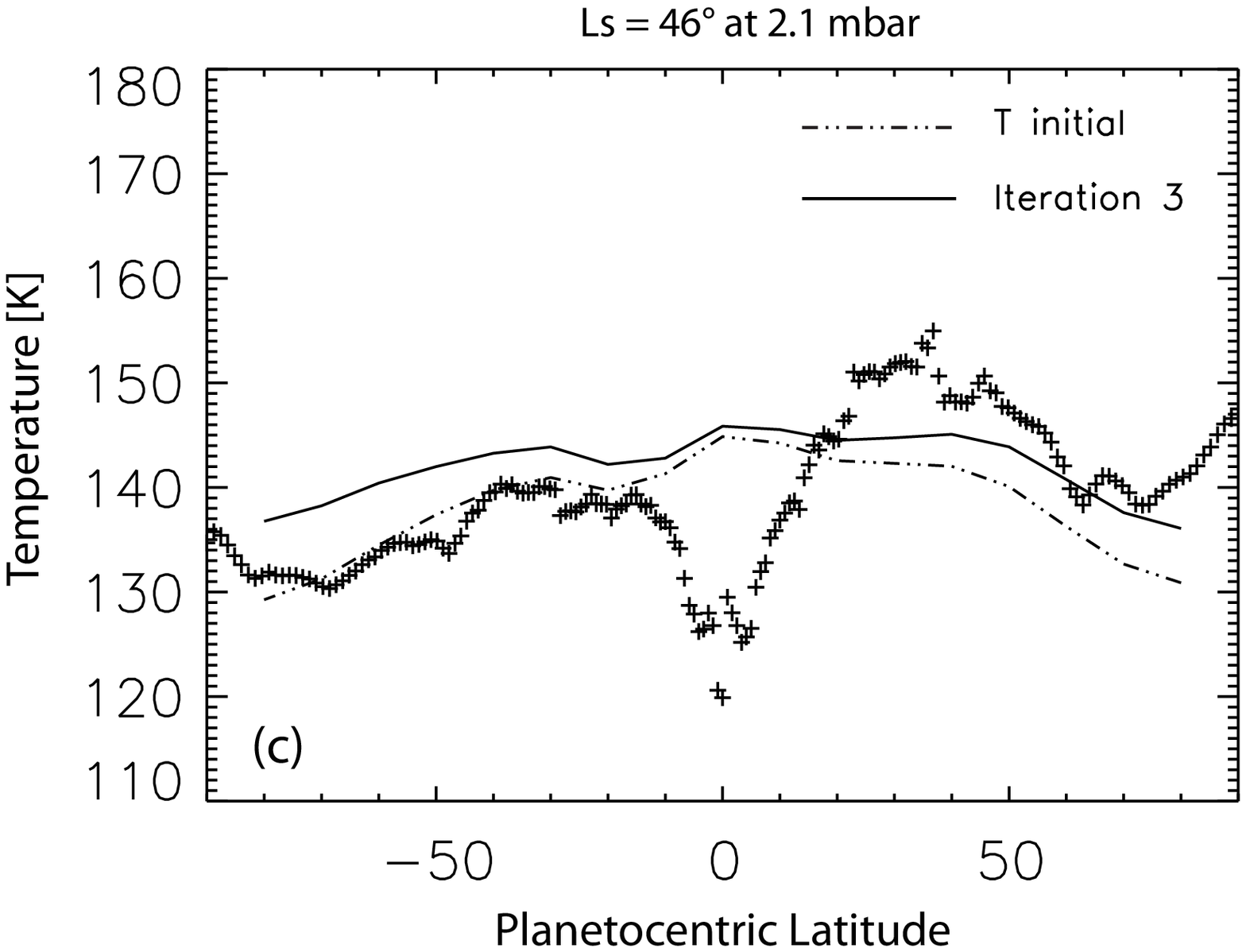}
      \caption{Temporal evolution of the meridional profile of temperature at 2.1\,mbar, in 2005-2006 (top panel), 2010-2011 (middle panel) and 2013-2014 (lower panel) observed by Cassini \citep{Fletcher2015,Cassini2015}. Prediction of the meridional temperature profile assuming the initial distribution of absorbers and coolants and at iteration 3 are denoted by dashed-triple dotted and solid lines, respectively. The errors bar on the temperature have not been overplotted, for the sake of clarity. The typical errors on temperature are $\pm$ 2-3\,K. 
      }
      \label{fig:Comp_Temp_nadir}
   \end{figure}

Nadir observations of temperature provide a good monitoring of the spatial and temporal evolution of Saturn's stratospheric temperature at 2.1\,mbar. The evolution of the temperature at this pressure level is presented in Fig. \ref{fig:Comp_Temp_nadir}. The observed temperatures were averaged into 1 year-long temporal bins, i.e. 2005-2006 (L$_s$ $\approx$ 300$^{\circ}$-313$^{\circ}$, Fig  \ref{fig:Comp_Temp_nadir}a); 2010-2011 (L$_s$ $\approx$ 5$^{\circ}$-17$^{\circ}$, Fig  \ref{fig:Comp_Temp_nadir}b) and 2013-2014 (L$_s$ $\approx$ 41$^{\circ}$-52$^{\circ}$, Fig  \ref{fig:Comp_Temp_nadir}c). The meridional profile of the initial thermal field and the converged thermal field are presented over the same period, i.e. L$_s$ = 306$^{\circ}$ for the 300$^{\circ}$-313$^{\circ}$ period, L$_s$ = 11$^{\circ}$ for the 5$^{\circ}$-17$^{\circ}$ period and L$_s$ = 46$^{\circ}$ for the 41$^{\circ}$-52$^{\circ}$ period.

The equatorial temperatures predicted by the initial thermal field and the converged one are in good agreement with one another. Indeed, as shown in Figs. \ref{fig:Results_1} and \ref{fig:Results_3}, the higher the latitude, the greater the feedback between the chemical composition and the thermal structure. Due to the low feedback as well as the good reproduction of the a priori abundance values in the equatorial region used to compute the initial thermal field, the predicted temperatures at iteration 3 are close to the initial ones. However, the predicted thermal fields, both the initial and converged, cannot explain the temperature peak in the equatorial region observed by Cassini in 2005-2006 (Fig. \ref{fig:Comp_Temp_nadir}a) as well as the temperature minimum observed in 2013-2014 (Fig. \ref{fig:Comp_Temp_nadir}c). These local extrema in the temperature field are associated with the Saturn's Semi-Annual Oscillation (SSAO) \citep{Fouchet2008, Orton2008, Guerlet2011}. We remind the reader that neither the photochemical nor the radiative seasonal model forreproduce this phenomenon which can only be studied in the framework of a 3D-GCM.

The local minima predicted in 2005-2006 (Fig. \ref{fig:Comp_Temp_nadir}a) between 20$^{\circ}$N and 30$^{\circ}$N are caused by the ring occultation (see for instance \citet{Greathouse2008} and \citet{Hue2015}). Because of the thermal inertia at this pressure level, these effects are still observed in 2010-2011 (Fig. \ref{fig:Comp_Temp_nadir}b) in the northern hemisphere, although the ring shade is on the southern hemisphere at that time of the year. Finally, in 2013-2014 (Fig. \ref{fig:Comp_Temp_nadir}c), these effects are observed in the southern hemisphere, between 20$^{\circ}$S and 30$^{\circ}$S. These local minimums, linked to the ring shadowing, are consistent with the radiative model of \citet{Guerlet2014}, but disagree with the observations, which never show such a feature under the ring. A dynamical scenario involving heating associated with subsidence under the ring's shadow was previously proposed by \citet{Guerlet2009,Guerlet2010}, and is consistent with the predicted circulation of \citet{Friedson2012}.

Fig. \ref{fig:Comp_Temp_nadir} suggests that accounting for the feedback between the chemical composition and the atmospheric temperature does not improve the fit of the absolute value of the temperature observed by Cassini. However, accounting for this feedback modifies the meridional gradients of the predicted temperatures.

The meridional gradients of temperature of the initial thermal field, the converged thermal field and the Cassini/CIRS observations are summarized in Table \ref{fig:tableau_gradient_T_nadir}. The latitudinal range used to calculate these gradients was adjusted in order to remove undesirable effects on both observed and predicted temperatures. Latitudes below $\pm$ 20$^{\circ}$ were not accounted for in order to remove the effects from SSAO on the observed temperatures. Such effects result from a wave-driven interaction and cannot be reproduced in non-GCM models. Low- to mid-latitudes were occasionally excluded in order to remove signature from the ring shadowing on the predicted thermal field, which was not observed by Cassini (e.g. Fig.  \ref{fig:Comp_Temp_nadir}a and \ref{fig:Comp_Temp_nadir}b from 20$^{\circ}$N-40$^{\circ}$N or \ref{fig:Comp_Temp_nadir}c from 10$^{\circ}$S-30$^{\circ}$S).

\renewcommand{\arraystretch}{1.5}
\begin{table*}[t]
\footnotesize
\begin{center}
\begin{tabular}{ c c c c c }
\hline
Observation & Hemisphere & Temperature & Temperature  & Temperature   \\
 period &  & gradient [K/$^{\circ}$] & gradient [K/$^{\circ}$] &  gradient [K/$^{\circ}$] \\
  &  & (observed) &  (initial) &  (converged) \\
\hline
\hline
2005-2006 & South (20$^{\circ}$S-80$^{\circ}$S) & -0.349 \,$\pm$\, 0.040 & -0.136 & -0.241 \\
(Fig. \ref{fig:Comp_Temp_nadir}a)& North (40$^{\circ}$N-80$^{\circ}$N) & -0.140 \,$\pm$\, 0.015 & -0.296 & -0.210 \\
\hline
2010-2011 & South (20$^{\circ}$S-80$^{\circ}$S) & -0.072 \,$\pm$\, 0.006 & 0.093 & -0.007 \\
(Fig. \ref{fig:Comp_Temp_nadir}b)& North (40$^{\circ}$N-80$^{\circ}$N) & -0.429 \,$\pm$\, 0.063 & -0.543 & -0.472 \\
\hline
2013-2014 & South (30$^{\circ}$S-80$^{\circ}$S) & 0.211 \,$\pm$\, 0.027 & 0.248 & 0.149 \\
(Fig. \ref{fig:Comp_Temp_nadir}c)& North (30$^{\circ}$N-80$^{\circ}$N) & -0.316 \,$\pm$\, 0.025 & -0.254 & -0.197 \\
\hline
\end{tabular}
\end{center}
\caption{Comparison between the meridional gradients of temperature at 2.1\,mbar of Fig. \ref{fig:Comp_Temp_nadir}. The gradients are given for the initial thermal field, the converged thermal field and the Cassini/CIRS observations \citep{Fletcher2015,Cassini2015}.}
\label{fig:tableau_gradient_T_nadir}
\normalsize
\end{table*}
\renewcommand{\arraystretch}{3.0}

\normalsize

The southern meridional gradient of the predicted temperature for the 2005-2006 period (Fig. \ref{fig:Comp_Temp_nadir}a) is increased at iteration 3 with respect to the initial one. This gradient (-0.241 K/$^{\circ}$) is significantly closer to the value observed (-0.349 \,$\pm$\, 0.040 K/$^{\circ}$) than the one predicted assuming a uniform distribution of absorbers and coolants (-0.136 K/$^{\circ}$). At the same period in the northern hemisphere, the meridional gradient at iteration 3 is slightly flattened with respect to the initial one. This converged thermal field gradient (-0.210 K/$^{\circ}$) also represents an improvement with respect to the initial one (-0.296 K/$^{\circ}$), when compared to Cassini observations (-0.140\,$\pm$\, 0.015 K/$^{\circ}$).

In the 2010-2011 period, the temperature gradients are better reproduced with the converged field both in the southern and in the northern hemisphere (-0.007 K/$^{\circ}$ and -0.472 K/$^{\circ}$ respectively) with respect to the initial thermal field (0.093 K/$^{\circ}$ and -0.543 K/$^{\circ}$) when compared to the Cassini observations (-0.072 \,$\pm$\, 0.006 K/$^{\circ}$ and -0.429 \,$\pm$\, 0.063 K/$^{\circ}$).


In 2013-2014 (Fig. \ref{fig:Comp_Temp_nadir}c), the temperatures predicted by the converged thermal field both in the southern and the northern hemisphere (0.149 K/$^{\circ}$ and -0.197 K/$^{\circ}$) are in worse agreement with the Cassini observations (0.211 \,$\pm$\, 0.027 K/$^{\circ}$ and -0.316 \,$\pm$\, 0.025 K/$^{\circ}$) with respect to the initial thermal field (0.248 K/$^{\circ}$ and -0.254 K/$^{\circ}$), in terms of meridional gradient of temperature.




Although accounting for the feedback between the thermal field and the chemical composition does not provide a better overall fit of the temperature absolute value measured by Cassini, it does bring the meridional gradients of Temperature closer to the ones observed by Cassini in most cases ($\mathtt{\sim}$ 67\%).


\subsection{Comparison with limb-observations}
\label{sss:Limb_Observations}

\begin{figure}[htp]
   \centering
   \includegraphics[width=0.55\columnwidth]{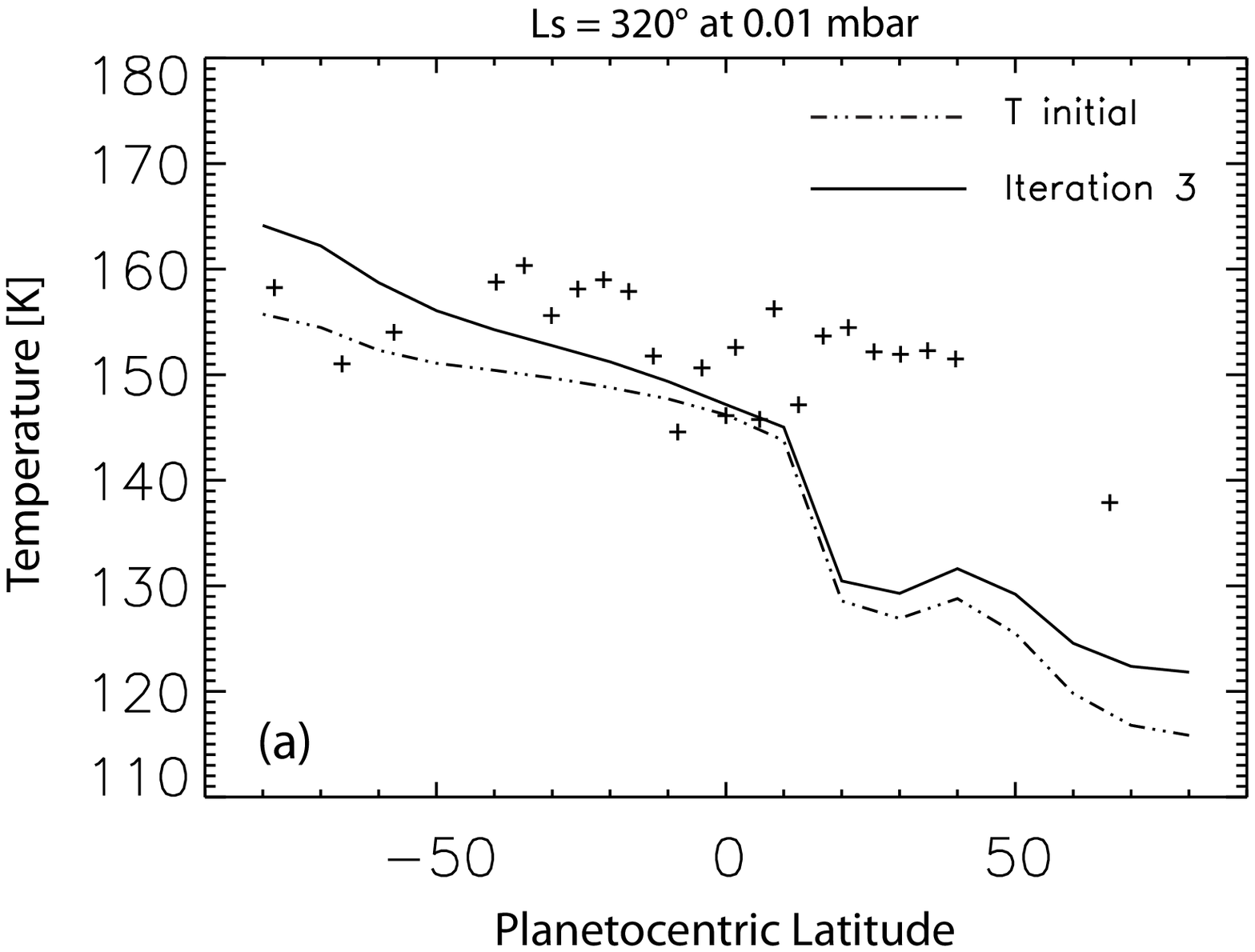}\vspace{0.3cm}\\
   \includegraphics[width=0.55\columnwidth]{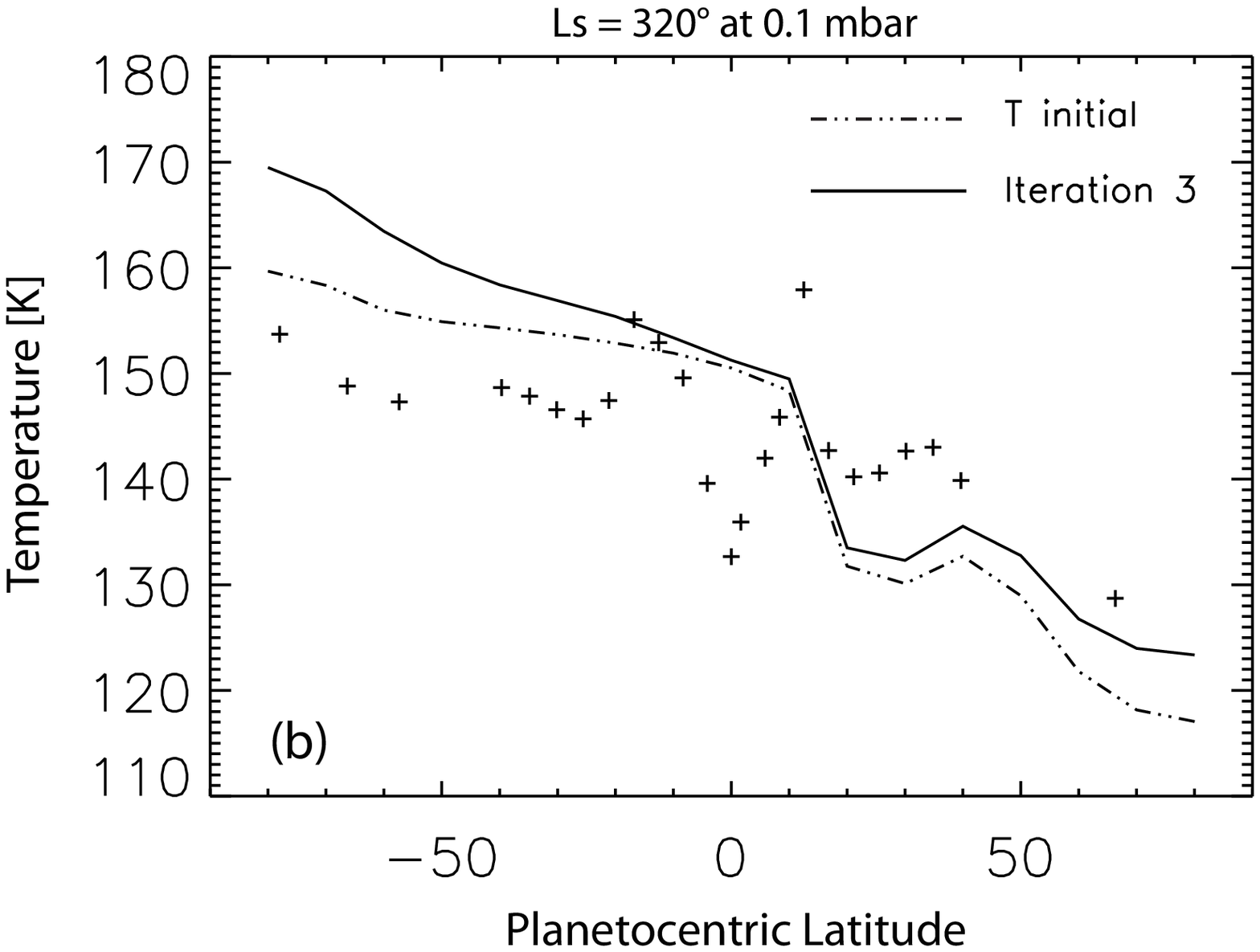}\vspace{0.3cm}\\
   \includegraphics[width=0.55\columnwidth]{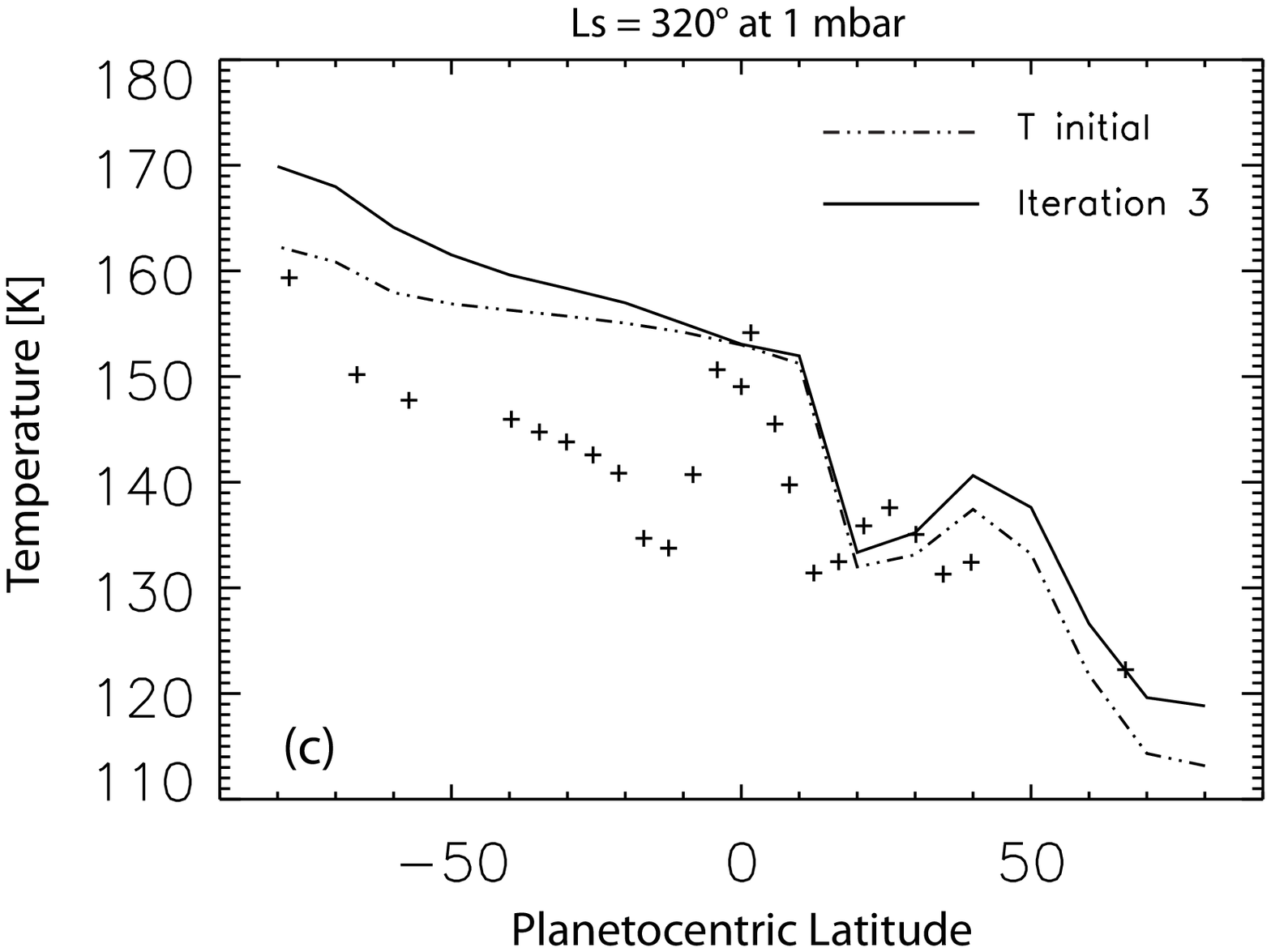}
      \caption{Comparison between retrieved temperature of \citet{Guerlet2009} and predicted thermal field at L$_s$ = 320$^{\circ}$. Dashed triple-dotted lines and solid lines assume repectively the initial and converged distribution of absorbers and coolants. Upper, middle and lower panel denote the evolution of these stratospheric temperatures at 10$^{-2}$, 10$^{-1}$ and 1\,mbar, respectively. The typical uncertainties on temperature are $\pm$ 1-2\,K}
      \label{fig:Comp_Temp_Limb}
   \end{figure}

Despite the lower temporal coverage of the Cassini/CIRS limb-observations, these data provide valuable information on the vertical structure of the atmospheric emission and therefore serve as good vertical constrains of the stratospheric temperatures and strongest emitting hydrocarbons. \citet{Guerlet2009,Guerlet2010} retrieved the vertical structure of the temperature, as well as the C$_2$H$_2$ and C$_2$H$_6$ adundances, from roughly 5\,mbar to 5$\times$ 10$^{-3}$\,mbar and over a temporal window ranging from L$_s$ = 300$^{\circ}$ to 340$^{\circ}$. We present here a comparison between the initial thermal field as well as the converged thermal field with the observations of the temperature on Fig. \ref{fig:Comp_Temp_Limb}, at the pressure levels sounded by CIRS limb observations.

Similarly to the comparison with Cassini nadir observations, the converged thermal field at L$_S$ = 320$^{\circ}$ generally suggests a modification of the meridional thermal gradient. Because the equatorial zone is altered by the SSAO and because there is a lack of data in the northern hemisphere, we discuss only the predicted meridional gradients in the southern hemisphere. The table \ref{fig:tableau_gradient_T_Limb} summarized the observed and predicted meridional temperature gradients.

At a pressure level of 10$^{-2}$\,mbar (Fig. \ref{fig:Comp_Temp_Limb}a), the Cassini observations tend to suggest a small meridional thermal gradient in the southern hemisphere (0.024 \,$\pm$\,0.023\,K/$^{\circ}$). Despite the large error bars, this gradient seems to be in better agreement with the initial thermal field (-0.111 K/$^{\circ}$) than with the converged one (-0.212 K/$^{\circ}$). The situation is also true at 0.1\,mbar where the converged thermal field (-0.243 K/$^{\circ}$) is in worse agreement with the observed one (-0.092\,$\pm$\,0.027\,K/$^{\circ}$) than the initial field (-0.112 K/$^{\circ}$). Finally, at 1\,mbar (Fig. \ref{fig:Comp_Temp_Limb}b), the predicted converged temperature gradient (-0.223 K/$^{\circ}$) is closer to the observed one (-0.265\,$\pm$\,0.027\, K/$^{\circ}$) than the initial one (-0.120 K/$^{\circ}$).


\renewcommand{\arraystretch}{1.5}
\begin{table*}[t]
\footnotesize
\begin{center}
\begin{tabular}{ c c c c c }
\hline
 Pressure & Hemisphere & Temperature & Temperature  & Temperature  \\
 level &  & gradient [K/$^{\circ}$] & gradient [K/$^{\circ}$] & gradient [K/$^{\circ}$] \\
  &  & (observed) & (initial) &  (converged) \\
\hline
\hline
0.01\,mbar & South (20$^{\circ}$S-80$^{\circ}$S) & 0.024 \,$\pm$\, 0.023 & -0.111 & -0.212 \\
(Fig. \ref{fig:Comp_Temp_Limb}a)&  &  &  &  \\
\hline
0.1\,mbar & South (20$^{\circ}$S-80$^{\circ}$S) & -0.092  \,$\pm$\, 0.027 & -0.112 & -0.243 \\
(Fig. \ref{fig:Comp_Temp_Limb}b)&  & &  &  \\
\hline
1\,mbar & South (20$^{\circ}$S-80$^{\circ}$S) & -0.265   \,$\pm$\,  0.027 & -0.120 & -0.223 \\
(Fig. \ref{fig:Comp_Temp_Limb}c)&  &  &  &  \\
\hline
\end{tabular}
\end{center}
\caption{Comparison between temperature meridional gradients at different pressure levels (0.01\,mbar, 0.1\,mbar and 1\,mbar) in the southern hemisphere. The gradients are given for the initial thermal field, the converged thermal field and the Cassini/CIRS observations \citep{Guerlet2009}.}
\label{fig:tableau_gradient_T_Limb}
\normalsize
\end{table*}
\renewcommand{\arraystretch}{3.0}

\normalsize

\subsection{Discussions}
\label{sss:Discussion}

As noted above, accounting for the feedback between the chemical composition and the thermal structure does not represent an improvement when comparing the predicted absolute value of the temperature and with the Cassini observations. Indeed, the temperatures predicted by the converged thermal field are generally warmer than the observed values by up to 10\,K while the initial thermal field most of the time predicts temperature closer from the observed value. However, accounting for this feedback provides a better fit of the observed meridional gradients of temperature. We remind the reader that neither the photochemical model nor the radiative seasonal model account for the atmospheric dynamics. Dynamics are expected to modify the distribution of absorbers and coolants and therefore impact the thermal field. Therefore, it seems that only a 3D-GCM that simultaneously accounts for the hydrocarbon chemistry produced by the methane photolysis would be trusworthy in the atmospheric temperature predictions. Furthermore, the temperature perturbations caused by the SSAO could only be investigated within the framework of such a model.

The few existing 3D-GCMs of Saturn \citep{Friedson2012, Spiga2014, Spiga2015} do not account for the hydrocarbon chemistry, as they assume an observed vertical profile of C$_2$H$_2$ and C$_2$H$_6$, and a predicted vertical profile of CH$_4$. These profiles are assumed to be meridionally uniform and without any temporal evolution.

At the moment, the present coupling between the photochemical model and the radiative seasonal model represents the most reliable description of the feedback between the chemical composition and the thermal structure in Saturn, despite the fact that it does not yet include the atmospheric dynamics. Such a coupling should be investigated in future 3D-GCM.


\section{Conclusions}
\label{s:Conclusion}

In this work, we have coupled a radiative seasonal model \citep{Greathouse2008} to a photochemical model \citep{Hue2015} in order to study Saturn's stratospheric composition and temperature. The former one gives a prediction of the thermal field in Saturn's stratosphere given a priori hydrocarbon distributions while the latter one predicts these hydrocarbon distributions, given a pressure-temperature background. These hydrocarbons are crucial as they rule Saturn's atmospheric heating and cooling rates and therefore its temperature \citep{Bezard1985, Greathouse2008, Guerlet2014}. This temperature and its seasonal evolution are also important because it has been shown that they influence the atmospheric chemistry \citep{Hue2015}.

In the present work, both the photochemical and the radiative seasonal models account for the evolution of the seasonally variable parameters such as obliquity, eccentricity of Saturn's orbit and ring shadowing. The radiative seasonal model was first used to compute the thermal field, using Cassini limb observations of \citet{Guerlet2009} as a priori distributions of absorbers and coolants. The computed thermal field was then used as the pressure-temperature background in the photochemical model, similarly to the previous work of \citet{Hue2015}, in which the results were compared to Cassini observations \citep{Guerlet2009}. The abundance field computed by the photochemical model was then injected into the radiative seasonal model, to compute an updated thermal field. Several iterations between the radiative seasonal model and the photochemical model were needed for the thermal field and the abundance field to converge.


This study suggests two major new insights:

$\bullet$ The feedback between the radiative seasonal model and the photochemical model predicts that the moment of the year when the temperature peaks at high latitudes is affected. Indeed, when using the meridionally uniform spatial distribution of absorbers and coolants, this peak to expected to occur at the solstice itself. However, when accounting for the seasonal evolution of absorbers and coolants, this peak is shifted before the solstice, i.e. during the spring season. As an example, at 80$^{\circ}$ in both hemisphere, the temperature peak at 10$^{-2}$\,mbar is happening half a season (3-4 Earth years) earlier than previously predicted by radiative seasonal model that uses uniform distribution of coolants. The main stratospheric coolants in Saturn atmosphere are C$_2$H$_2$ and C$_2$H$_6$. At the beginning of the spring season at high latitudes, which just experienced winter, the amount of these coolants is low due to the previous winter's low insolation level. Therefore, the amount of these coolants is not high enough to efficiently cool the stratosphere, which causes this early temperature peak. By the middle of the spring season, the amount of coolants becomes high enough so that the stratosphere cools slowly all along the summer season. This effect is particularly pronounced at low-pressure levels, where the C$_2$H$_2$ and C$_2$H$_6$ chemical timescales are short, and progressively disappears at higher-pressure levels. On the other hand, the thermal field has a small feedback on the abundance distributions. At 10$^{-2}$\,mbar, this feedback never exceeds 11\% and 9\% for C$_2$H$_2$ and C$_2$H$_6$.

$\bullet$ Accounting for that feedback modifies the predicted meridional gradients of temperature. Although the absolute values of the temperature predicted by the converged thermal field are generally in worst agreement with Cassini observations with respect to the initial thermal field, the predicted meridional gradients of temperature around the millibar pressure level are, in most of the cases, closer to the observed one. At lower pressure levels, the meridional gradients of temperature are in worst agreement with Cassini observations suggesting that Saturn's upper stratosphere departs from pure radiative equilibrium as previously suggested by \citet{Guerlet2014}.

It is remarkable to note that very little a priori information from observations were used to compute the thermal and abundance converged field. In fact, the only poorly constrained parameter used in the photochemical model is the eddy diffusion coefficient that rules how fast the methane photolysis by-products diffuse down to the lower stratosphere. This eddy diffusion coefficient is taken constant with latitude and seasons, due to the lack of constraints on that parameter and consistent with \citet{Hue2015}.

Finally, this work underlines the importance of coupling the chemistry with the radiative calculations, as demonstrated by the sensitivity of the feedback that the amount of coolant has on the predicted temperature. Despite the moderate agreement with Cassini data we found in terms of temperature absolute values, this is nevertheless the approach to follow toward a better comprehension of the seasonal evolution of the giant planet stratospheric thermal field as well as the potential radiatively-driven circulations.

\section{Acknowledgement}

      This work has been supported by the \textit{Investissements d'avenir} program from \textit{Agence Nationale de la Recherche} under the reference ANR-10-IDEX-03-02 (IdEx Bordeaux). Part of this work was done by V.H. at Southwest Research Institute, San Antonio. T.G, working at Southwest Research Institute was supported by NASA PATM grant NNX13AHIG.

\bibliographystyle{elsarticle-harv}

\bibliography{aa}


\end{document}